\documentclass[aps,pre,showpacs,onecolumn]{revtex4-1}
\usepackage{graphicx}
\usepackage{dcolumn}
\usepackage{amssymb,amsmath}
\usepackage{natbib}
\usepackage{subfigure}
\usepackage{epstopdf}
\usepackage{placeins}
\usepackage{color}
\usepackage{epstopdf}

\begin{document}

\title{Large Eddy Simulation studies of the effects of alignment and wind farm length}
\author{Richard J. A. M. Stevens$^{1,2}$ Dennice F. Gayme$^{1}$ and Charles Meneveau}
\affiliation{
$^1$Department of Mechanical Engineering $\&$ Center for Environmental and Applied Fluid Mechanics, Johns Hopkins University, Baltimore, Maryland 21218, USA\\
$^2$Department of Physics, Mesa+ Institute, and J.\ M.\ Burgers Centre for Fluid Dynamics, University of Twente, 7500 AE Enschede, The Netherlands}

\date{\today}

\begin{abstract}
Large eddy simulations of wind farms are performed to study the effects of wind turbine row alignment with respect to the incoming flow direction. Various wind farms with fixed stream-wise spacing (7.85 rotor diameters) and varying lateral displacements and span-wise turbine spacings are considered, for a fixed inflow direction. Simulations show that, contrary to common belief, a perfectly staggered (checker-board) configuration does not necessarily give the highest average power output. Instead, the highest mean wind farm power output is found to depend on several factors, the most important one being the alignment that leads to minimization of wake effects from turbines in several upstream rows. This alignment typically occurs at significantly smaller angles than those corresponding to perfect staggering. The observed trends have implications for wind farm designs, especially in sites with a well-defined prevailing wind direction.
\end{abstract}


\maketitle

\section{Introduction}
Wind energy contributed about $3\%$ to the global electricity production at the end of 2011 \cite{WWER2011} and this contribution is expected to increase to about $20 \%$ by $2030$ \cite{eu2007,us2008}. High levels of wind penetration have already been achieved in several countries in Europe, such as Denmark ($26\%$), Portugal ($16\%$), and Spain ($16\%$). However, in order to realize global targets for wind power production more and larger wind farms will be required.

The effects of large wind farms at regional and global scales are often modeled by considering the wind farms as surface roughness elements or net drag coefficients. These surface elements increase the effective roughness length of the ground, which needs to be parameterized. Examples of this approach are found in studies that investigate the effect of large wind farms on the global climate \cite{kei04,wan10}, regional meteorology \cite{bai04}, or short time weather patterns \cite{bar10,zho12}. In these simulations the horizontal computational resolution is often significantly coarser than the height of the Atmospheric Boundary Layer (ABL) and therefore insufficient to capture important physical mechanisms.

Details of wind farm ABL interactions can be modeled with Large Eddy Simulations (LES). In the past, simulations of interactions between the ABL and wind farms have often been done using the Reynolds-Averaged Navier-Stokes (RANS) approach \cite{san11,ver03}, in order to make the computations affordable. Now however, LES is becoming more affordable \cite{jim08,jim10,cal10,iva10,san11,wu11,chu12} and by resolving parts of the spectrum of turbulence, LES captures important features of unsteady, anisotropic turbulence in turbine wakes interacting with the ABL. 

\begin{figure}
\begin{center}
\subfigure{\includegraphics[width=0.98\textwidth]{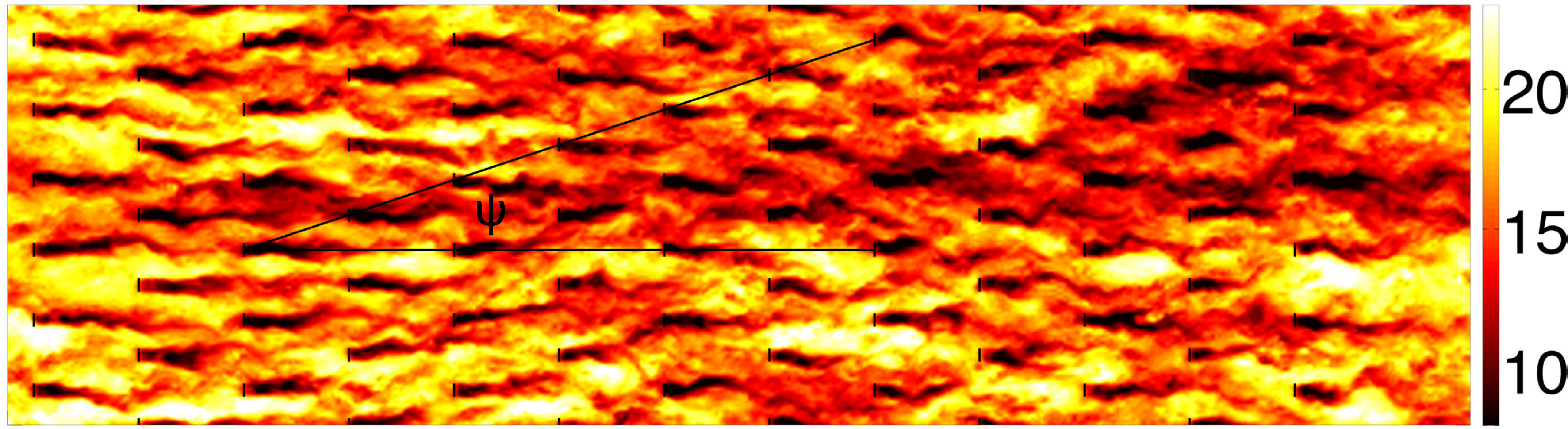}}
\caption{Snapshot of the stream-wise velocity at hub-height in a staggered wind farm with stream-wise and span-wise spacings of $S_x=7.85D$ and $S_y=5.23D$, respectively (case A3). The figure also shows the parameter $\psi$ that is used in this study to determine the wind farm layout. This angle $\psi=\arctan{\frac{S_{dy}}{S_x}}$ is defined with respect to the incoming flow direction, where $S_{dy}$ indicates the span-wise offset of subsequent turbine rows and $S_x$ is the stream-wise distance between the rows. The color scale indicates $u/u_{*}$, the stream-wise wind velocity in units of friction velocity $u_* = \sqrt{-L_z \partial_1 P/\rho}$, where $\partial_1 P$ is the imposed pressure gradient driving the flow in the precursor simulation.}
\label{figure1}
\end{center}
\end{figure}

There are several LES studies that model the interaction between one or two turbines and the ABL \cite{jim08,jim10,mo13,li13,mar12,cat12,tro10,tro11,sto12}, but only a limited number of LES investigations of full wind farms have been reported. Ivanell \cite{iva10} performed LES of two of the ten rows of the Horns Rev farm in Denmark and assumed periodic conditions in the span-wise direction to approximate the full plant aerodynamics. They pre-generated turbulent fields using Mann's synthetic turbulence method as turbulent inflow conditions \cite{iva10}. They then varied the wind inflow angle by $\pm 15^{\circ}$ with respect to the alignment of the turbine rows. Churchfield et al.\ \cite{chu12,chu12b,lee12} used a precursor LES of the ABL to create inflow conditions in their study of the Lillgrund wind farm. Their time-averaged power production for the aligned case agrees well with field observations up to the sixth turbine row. Meyers and Meneveau \cite{mey10} and Calaf et al.\ \cite{cal10,cal11} performed horizontally periodic LES to study infinitely long wind farms. They studied the effect of the spacing between the wind turbines on the total power output, as well as the scalar transport in the presence of wind farms. In agreement with wind tunnel experiments from Cal et al.\ \cite{cal10b}, they found that the vertical kinetic energy flux determines the average turbine power output in the fully developed regime of the wind farm. Wu and Port\'e-Agel \cite{wu11,wu13} showed with simulations of finite length wind farms that the relatively longer separation between consecutive downwind turbines in staggered wind farms as compared to aligned wind farms allows the wakes to recover more and exposes the turbines to higher local wind speeds. These simulations are in agreement with experiments from Chamorro et al.\ \cite{cha11b} which show a higher vertical momentum transfer into the hub-height plane for a staggered wind farm than for an aligned wind farm. Yang et al.\ \cite{yan12} showed with simulations that in infinitely large aligned wind farms the stream-wise spacing is more important for the average power production than the span-wise spacing. 

Engineering wind farm design tools have some difficulties predicting the deep array effects in large wind farms when standard parameters are used, because the interaction of turbulence generated by the wind turbines wakes with the overlying atmosphere is not fully taken into account \cite{bar09b}. LES can better model these large scale interactions and therefore provide useful reference points for trends observed in very large wind farms \cite{bar09b,wu13b}. Past work on small wind farms has shown significant differences in the power output of aligned and staggered wind farms, while relatively small differences, of the order of $5$ to $10\%$ have been observed in very large or infinite wind farms \cite{mey10,wu13}, and Archer et al.\ \cite{arc13b} investigated the sensitivity of the wind-wind power output to several design parameters using the Lillgrund wind-farm design as the reference-case. This raises questions about how turbine row alignment with respect to the wind inflow direction and overall wind farm size affect the power output. In this study we use LES to investigate the effect of the turbine alignment with respect to the inflow on the power production and the development of the wind farm boundary layer.

\section{Computational tool and simulation parameters}
In this paper, we model wind farms consisting of a regular array of wind turbines, each having a rotor diameter of $D=100$ m and a hub-height of $z_H=100$ m. The computational domain employed in our LES is $12.57 \times 3.14 \times 2$ km ($L_x\times L_y \times L_z$) in the stream-wise, span-wise and vertical directions, respectively.Due to the periodicity in the span-wise direction and the desire for turbines to be spaced uniformly in that direction the spacing is determined as $S_y = L_y/N_{sp}$, where $N_{sp}$ is the number of turbines in the span-wise direction and $L_y$ is the span-wise extent of the farm. In this work we study configurations with either $6$ or $9$ turbines in the span-wise direction, which leads to $S_y=5.23D$ and $S_y=3.49D$, respectively. Selecting $S_x=7.85$ as stream-wise spacing ensures $S_x/S_y=1.5$ and $S_x/S_y=2$, respectively. In order to study the influence of turbine alignment we change the wind farm layout by adjusting the angle $\psi=\arctan{\frac{S_{dy}}{S_x}}$ with respect to the incoming flow direction, where $S_{dy}$ indicates the span-wise offset from one turbine row to the next, see figure \ref{figure1}. The mean inflow remains in the x-direction in all cases. Thus $\psi=0$ degrees corresponds to an aligned wind farm. For the span-wise spacing of $5.23D$ a perfectly staggered (checker-board) arrangement corresponds to $\psi=\arctan[(5.23D/2)/7.85D]=18.43$ degrees and for the span-wise spacing of $3.49D$ this angle is $\psi=\arctan[(5.23D/2)/7.85D]=12.57$ degrees. Table \ref{table1} shows that the different alignment angles $\psi$ are distributed approximately uniformly between $\psi=0$ and the $\psi$ corresponding to the staggered case. Changing the wind farm layout by adjusting $\psi$ ensures that both the covered area and the number of turbines remains constant, which allows us to isolate the effect of the wind farm layout on the average power production.

We simulate the different finite length wind farms using a recently developed concurrent-precursor method to generate the inflow conditions \cite{ste14}. In the concurrent-precursor method we consider two computational domains simultaneously, i.e. in one domain we consider a turbulent ABL to generate the turbulent inflow conditions for the second domain in which the wind turbines are placed. In each domain we solve the filtered incompressible Navier-Stokes equations for neutral flows. In our code the nonlinear terms are evaluated in the rotational form. A pseudo-spectral discretization and thus double periodic boundary conditions are used in the horizontal directions, while centered second-order finite differencing is used in the vertical direction. The dynamic Lagrangian scale-dependent Smagorinsky model is used to calculate the subgrid-scale stresses \cite{bou05}. A second-order accurate Adams-Bashforth scheme is used for the time integration. The top boundary uses zero vertical velocity and a zero shear stress boundary condition. At the bottom surface a classic imposed wall stress \cite{moe84} boundary condition relates the wall stress to the velocity at the first grid point, with a prescribed roughness length-scale \cite{moe84}. A roughness length of $z_0=5\times10^{-5} L_z$ (where $L_z=2$ km) is used. Periodic boundary conditions are used in the span-wise direction. The flow is forced by using an imposed x-direction stream-wise pressure gradient $\partial_1 P$ in the periodic precursor simulation domain. As a result, in that domain one obtains a friction velocity given by $u_* = \sqrt{-L_z \partial_1 P/\rho}$ which will be used to non-dimensionalize velocity scales throughout the paper. We refer to Calaf et al.\cite{cal10} (appendix A) on how to relate this approach to velocity scales arising from geostrophic velocity forcing. We consider only neutral atmospheric conditions (no stratification), with a mean flow in the x-direction since a constant stream-wise pressure gradient is applied only in the x-direction of the precursor simulation. A caveat, which will be important when interpreting the results, is that under realistic conditions the incoming wind direction may change over various time-scales. 

\begin{table} 
\caption{Summary of the performed simulations. The columns from left to right indicate the name of the simulation case. The resolution in the stream-wise, span-wise, and vertical directions are denoted by $N_x$, $N_y$, and $N_z$, respectively,. The number of turbines in the stream-wise $N_{st}$ and span-wise $N_{sp}$ directions, the stream-wise $S_x$ and span-wise $S_y$ distance between the turbines, and a list of alignment angles, are also indicated.}
\label{table1}
\begin{center}
\begin{tabular}{|c|c|c|c| l |}
\hline
Case	& 	$N_x 	\times N_y 	\times N_z$	& $N_{st} \times N_{sp}$	& $S_x \times S_y$		& $\psi$ (degrees) \\	 \hline	
A1		&	$512		\times 64		\times 128$	& $13 \times 6$		 	& $7.85D \times 5.23D$	& $0.00; 1.91; 3.81; 5.71; 7.59; 9.46;$\\ 
 		&									&					&					& $11.31; 13.13; 14.93; 16.70; 18.43$ \\ \hline
A2		&	$768		\times 96		\times 192$	& $13 \times 6$			& $7.85D \times 5.23D$	& $0.00$ 	\\ \hline
A3		& $1024	\times 128		\times 256$		& $13 \times 6$			& $7.85D \times 5.23D$	& $0.00; 3.81; 7.59; 9.46; 11.31; 14.93; 18.43$ \\ \hline
A4		& $1536	\times 192		\times 384$		& $13 \times 6$			& $7.85D \times 5.23D$	& $0.00$ \\ \hline
B1		&	$512		\times 64		\times 128$	& $13 \times 9$			& $7.85D \times 3.49D$	& $0.00; 1.71; 1.91; 3.38; 5.08; 5.71;$ \\ 
 		&									&					&					& $6.77; 8.42; 9.46; 10.08; 11.31; 12.53$ 	 \\ \hline
B3		& $1024	 \times 128	\times 256$		& $12 \times 9$			& $7.85D \times 3.49D$	&$0.00; 3.38; 6.77; 9.46; 11.31; 12.53$ \\ \hline
 \end{tabular}
\end{center}
\end{table}

The turbines are modeled using an area averaged actuator disk method \cite{jim08,cal10,cal11,ste14}. The average power output of the turbines here is evaluated as equal to the mechanical energy loss in the fluid, namely $P= - \langle F U_d \rangle$, where $F=-\frac{1}{2}C_{T}^{'}\rho U_d^2 A$ is the local force used in the actuator disk model, $U_d$ is the disk averaged velocity, $A=\pi D^2/4$ is the turbine rotor area, $\rho$ is the density of the fluid, and $C_T^{'}=C_T/(1-a)^2$, where $a$ is the axial induction factor. Using typical values $C_T=4a(1-a)$ for $a=1/4$ leads to $C_T^{'}=4/3$ \cite{cal10,cal11,mey11,jim08,bur01}. The mechanical energy loss $P=\frac{1}{2} \rho C_T/(1-a)^2 \langle U_d^3\rangle A$, when equated to $\frac{1}{2} \rho C_P U_\infty^3 A$ and assuming $\langle U_d^3\rangle = (1-a)^3 U_\infty^3$ would correspond to $C_P = 0.56$ for $a=0.25$. I.e. to evaluate the dimensional power output with a realistic value of $C_P \approx 0.4$, the extracted power would be evaluated as $P= - \frac{0.4}{0.56} \langle F U_d \rangle$. However, since we are only interested in ratios of power, and we will only focus on ``region II'' operation, the precise implied value of $C_P$ is not relevant for this particular study. The $*3$ simulations are started from an initiation condition obtained from the lower $*1$ resolution cases run for $\approx20$ dimensionless time units (where the dimensionless time is in units of $H/u_*$). Subsequently the simulation is continued for $\approx1$ dimensionless time unit on the finer grid to ensure that the flow reaches a statistically stationary state on the new grid before data collection is started. Data is collected for roughly $6$ time units, which corresponds to $6\times2000/0.45/3600\approx7.5$ hours for $H=2000$m and $u_*=0.45$m/s.

\begin{figure}
\subfigure{\includegraphics[width=0.49\textwidth]{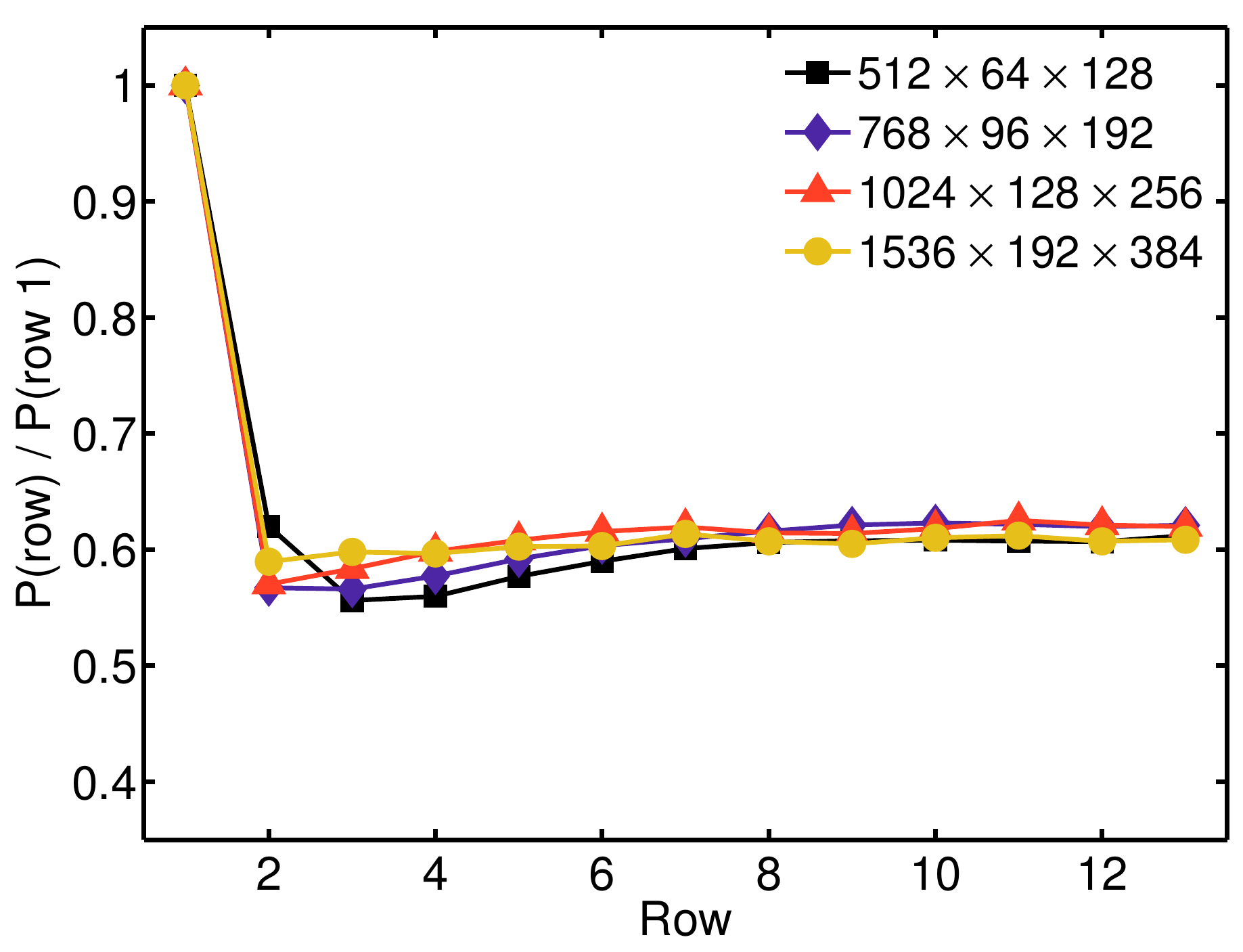}}
 \caption{The normalized average power output as function of the downstream position for different grid resolutions (cases A1-A4 for $\psi=0$, see table \ref{table1}) for the case of a stream-wise spacing of $7.85D$ and a span-wise spacing of $5.23D$. The normalization uses the power output of the first row.}
\label{figure2}
\end{figure}

In order to verify that the grid resolution is not influencing our main findings we performed simulations using different grid resolutions, ranging from $512 \times 64 \times 128$ to $1536 \times 192 \times 384$ grid points in the stream-wise, span-wise and vertical direction, respectively. Here, the number of points in the stream-wise direction is the combined number of grid points used in both computational domains. Figure \ref{figure2} shows the average power output ratio as function of the downstream position for an aligned wind farm using four different grid resolutions (cases A1-A4, see table \ref{table1}). The average turbine power output for a row is determined by calculating the time-average over the last Å6 dimensionless time units ($\approx7.5$ hours) of the simulation and by averaging over all turbines in that row. One can see that grid resolution has some influence on the average power output ratio, especially in the first couple of rows where the average power output can vary by a few of percent. However, the average power output in the fully developed regime of the wind farm appears to be independent of the grid resolution. For the purpose of this study this a reasonable accuracy as the effects and trends that we are interested in have a much more significant influence on the average power output. In addition, different modeling assumptions such as the turbine model, the corresponding turbine parameters, and the sub-grid scale model can also lead to some variation in the simulation output at varying resolutions. Previous work comparing the highest resolution data set to field measurement data, various other models and simulations of wind farms (see the inter comparison in Fig. 2 in Ref.\ \cite{ste13}), demonstrated that the trends seen in the LES results in figure \ref{figure2} show good agreement with the field measurement data of the Horns Rev wind farm. In order to verify that the grid resolution effects are consistent across a range of $\psi$ we performed several simulations on a $512\times64\times128$ and a $1024 \times 128 \times 256$ grid and found that the main trends were consistent on both grids. Therefore we only present the data obtained using the baseline $1024 \times 128 \times 256$ grid, unless stated otherwise. A detailed overview of the 38 different simulations performed and their parameters can be found in Table \ref{table1}.

\section{ LES Results on relative power output and alignment}
\begin{figure}
\subfigure[$\psi=0.00$ degrees]{\includegraphics[width=0.49\textwidth]{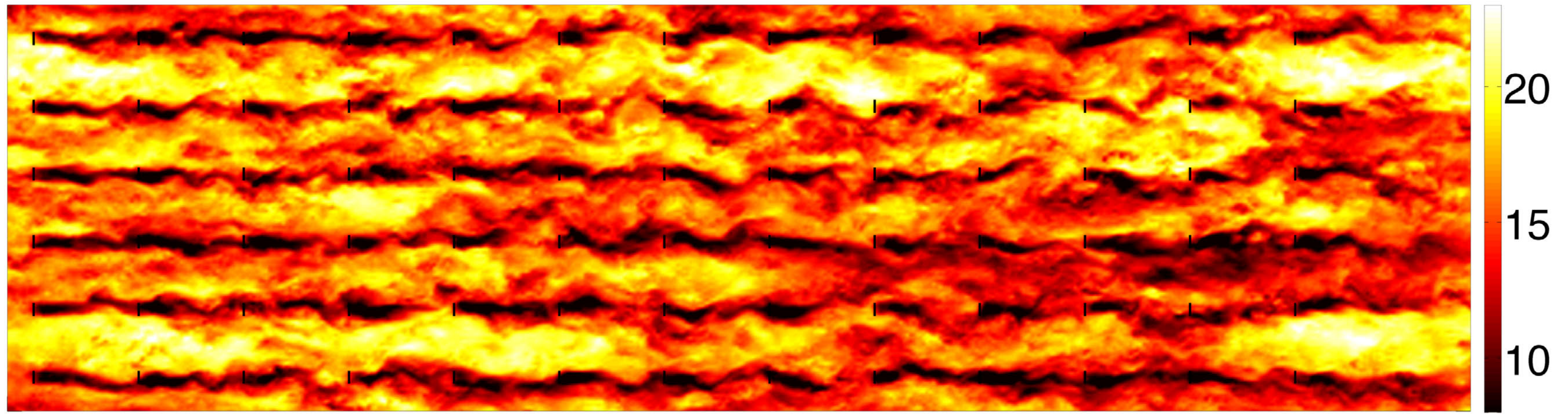}}
\subfigure[$\psi=18.43$ degrees]{\includegraphics[width=0.49\textwidth]{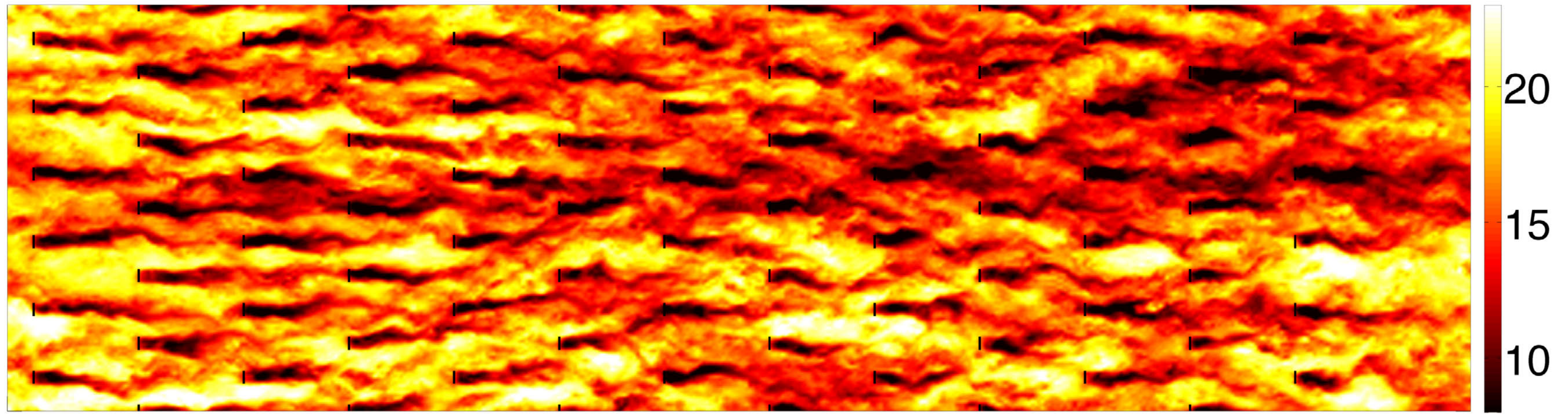}}
 \caption{Snapshot of the stream-wise velocity at hub-height for (a) aligned and (b) staggered wind farms with a stream-wise spacing of 7.85D and a span-wise of $5.23D$ (case A3).The color scale indicates $u/u_{*}$, the stream-wise wind velocity in units of friction velocity.}
\label{figure3}
\end{figure}

\begin{figure}
\subfigure[]{\includegraphics[width=0.49\textwidth]{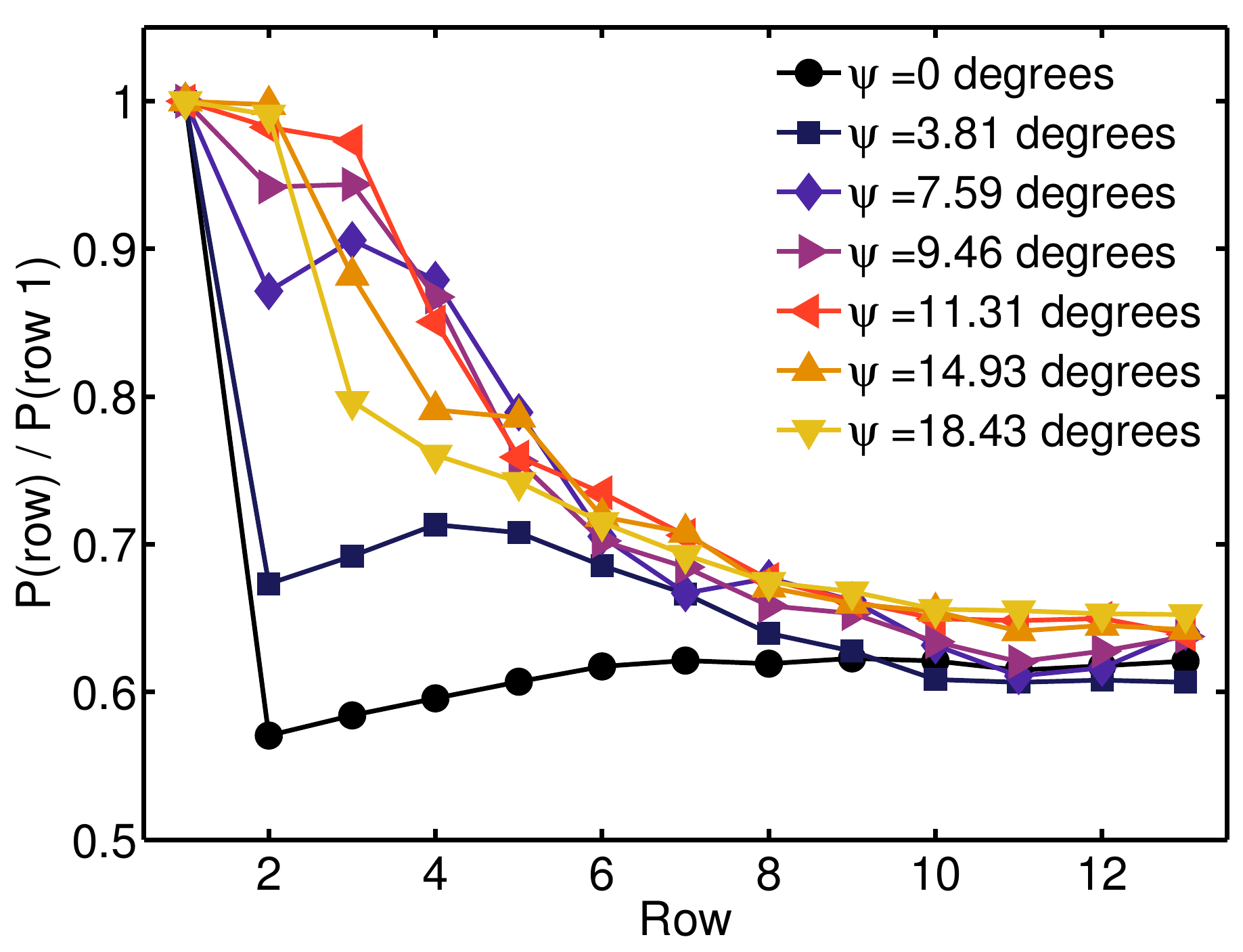}}
\subfigure[]{\includegraphics[width=0.49\textwidth]{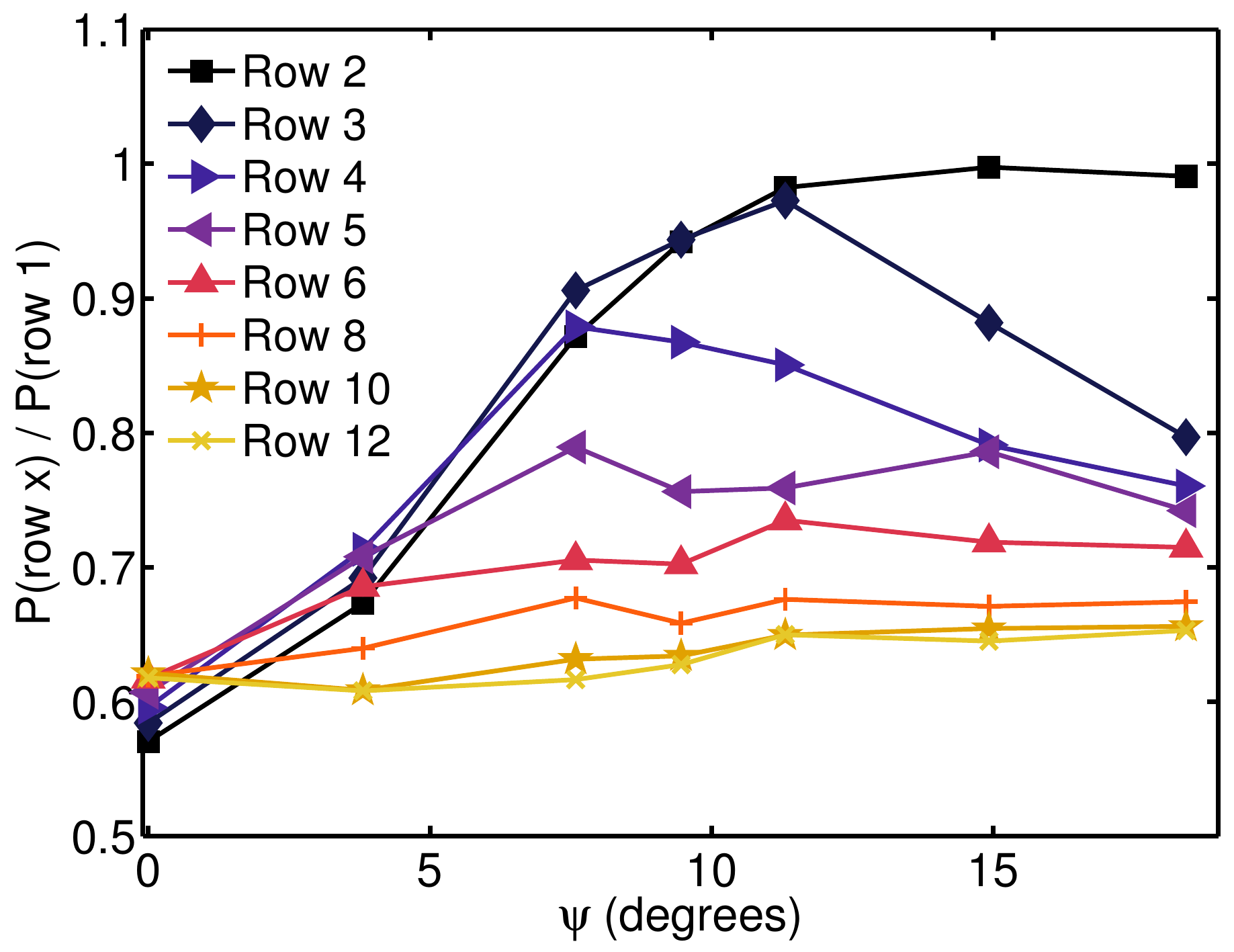}}
\subfigure[]{\includegraphics[width=0.49\textwidth]{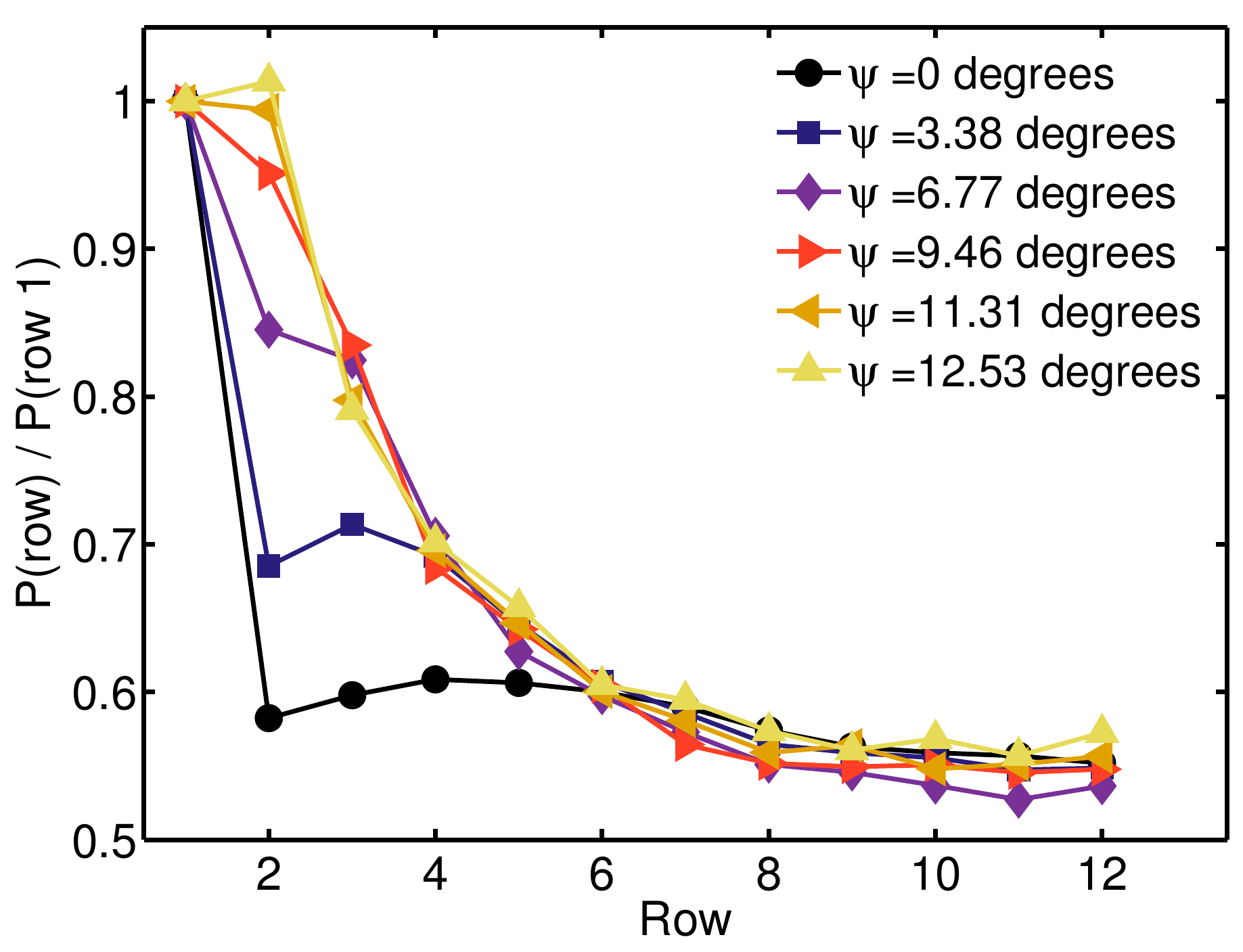}}
\subfigure[]{\includegraphics[width=0.49\textwidth]{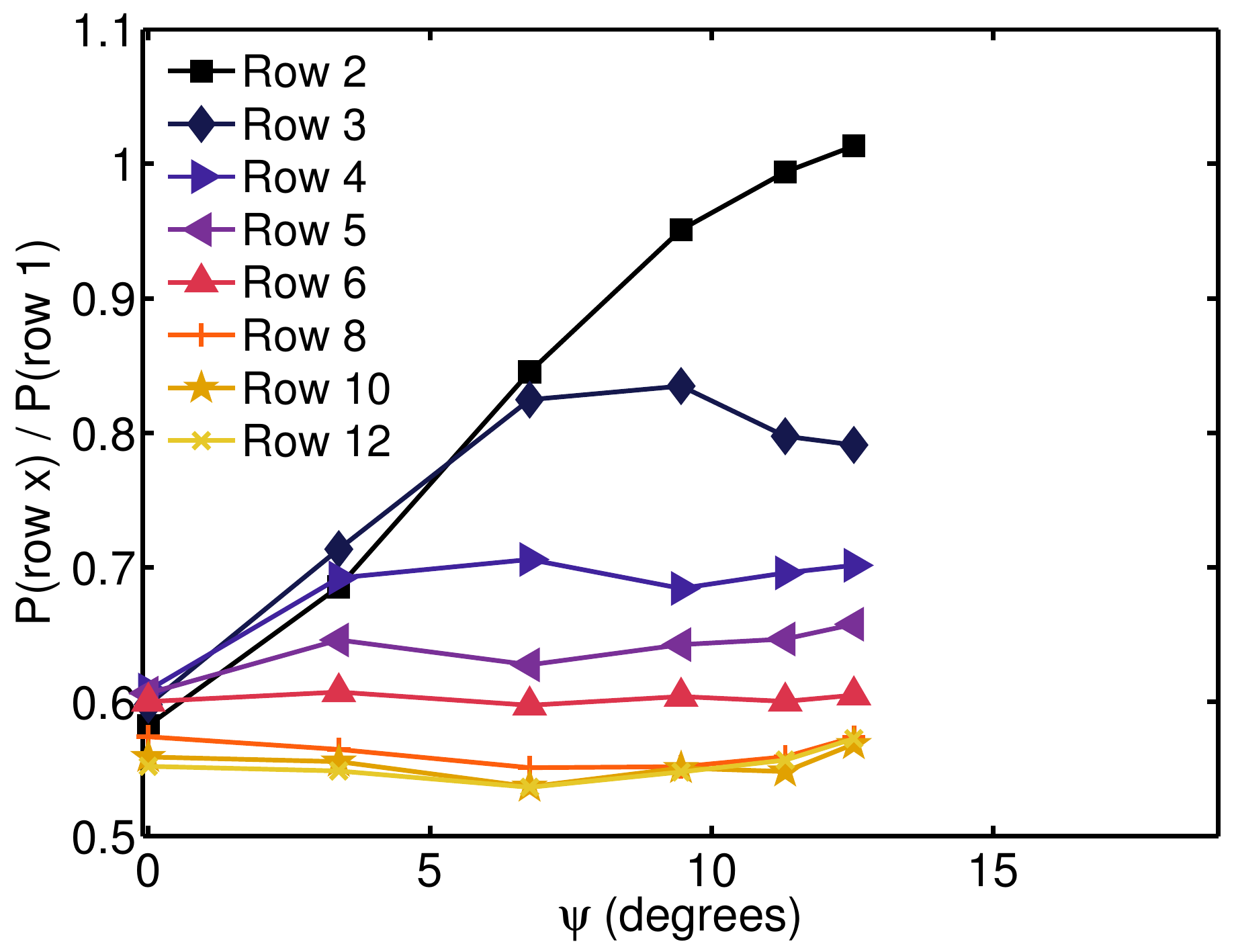}}
 \caption{The left panels (a) and (c) show the average power output as function of the downstream position for the different alignment angles $\psi$. The right panels (b) and (d) show the average power output for the different downstream turbine rows as function of $\psi$. The top (a,b) and bottom (c,d) panels respectively show the results for the wind farm with a span-wise spacing of $5.23D$ (case A3) and $3.49D$ (case B3).}
\label{figure4}
\end{figure}

The snapshots in figure \ref{figure3} show the stream-wise velocity (in units of the friction velocity $u_*$) at hub-height. Consistent with expectations, the velocity field in the wind farm is highly turbulent and the turbine wakes meander significantly. The figure reveals that high velocity patches of air can easily pass through the wind farm without encountering a turbine when an aligned configuration is used, which means that some of the incoming kinetic energy is not harvested even after passing along $13$ rows. In the staggered configuration, less high velocity wind patches seem to reach the end of the wind farm, which gives the impression that more energy is extracted from the flow. Figures \ref{figure4}a and \ref{figure4}c respectively show the normalized average power output ratio as function of the downstream position for the two different span-wise spacings ($5.23D$ and $3.49D$) and the different alignment angles $\psi$ described in table \ref{table1}. The figures reveal that for an aligned wind farm the average power production of turbines in the second row is about $40\%$ lower than in the first row and remains nearly constant for further downstream rows. This trend of sudden drop of power for aligned wind farms is well-known and has been documented in wind farms \cite{bar11} as well as wind tunnel studies \cite{wu13,new13}.

With increasing $\psi$ the power loss over the first couple of rows decreases more gradually than for the aligned configuration. A closer look at these figures reveals that for both span-wise spacings the highest average power output is obtained for $\psi \approx 11$ degrees. It is important to note that for the smaller span-wise distance of $3.49D$ this corresponds to a wind farm that is nearly staggered, i.e. at $12.57$ degrees, while this is an intermediate alignment when the span-wise spacing is $5.23D$ (whereas the staggered configuration corresponds to $18.43$ degrees).

\begin{figure}
\subfigure[]{\includegraphics[width=0.49\textwidth]{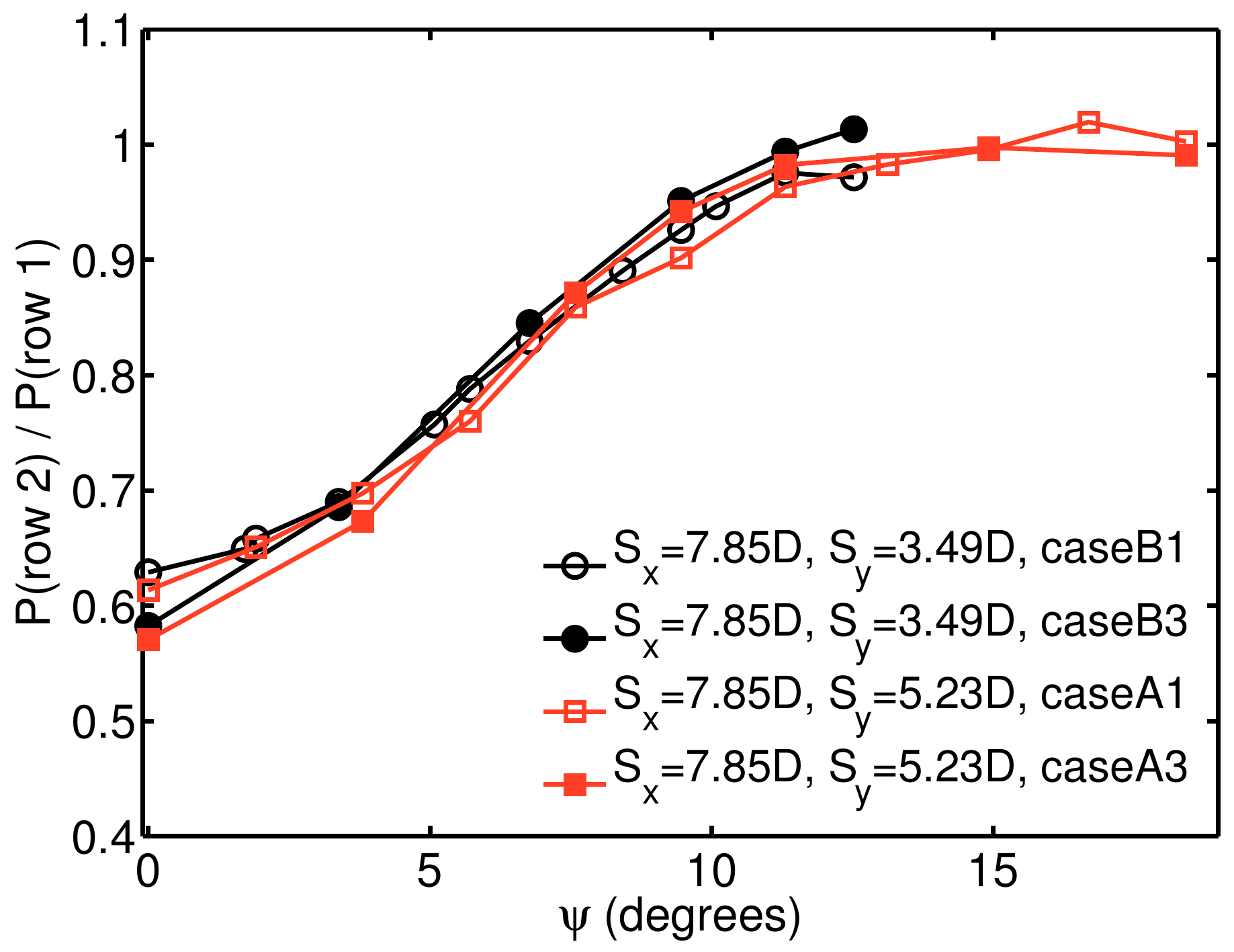}}
\subfigure[]{\includegraphics[width=0.49\textwidth]{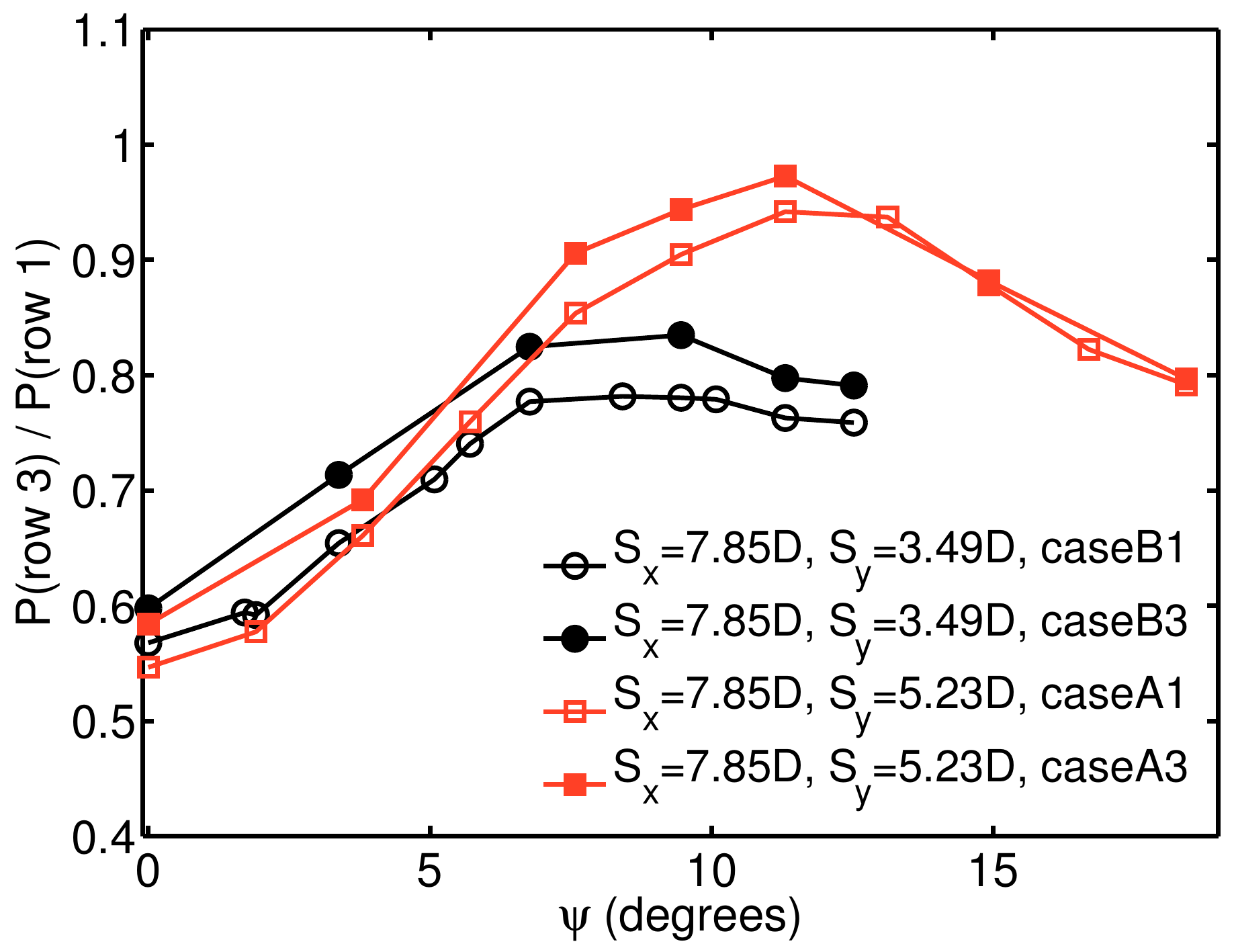}}
 \caption{Normalized averaged power output (power ratio) as function of the alignment angle $\psi$ for (a) the second and (b) the third row. Note that a similar trend is observed in the coarsest (cases $*1$) and finest (cases $*3$) resolution simulations. For a definition of the case numbers see table \ref{table1}.}
\label{figure5}
\end{figure}

In order to better understand this effect it is helpful to plot the results in a different way. Figure \ref{figure4}b and \ref{figure4}d show the normalized average power ratio as function of the alignment angle $\psi$ for the different turbine rows. For the second turbine row these figures reveal a significant power loss for the aligned or nearly aligned configurations, while the average power ratio at the second row approaches unity when $\psi \gtrsim 11$ degrees. The alignment becomes even more important for longer wind farms. Figure \ref{figure4}b shows that the average power ratio at the third turbine row as function of the alignment angle $\psi$ shows a maximum around $\psi \approx 11$ degrees when the span-wise spacing is $5.23D$. The average power ratio of the third turbine row is approximately equal to the average power ratio of the second row when $0 < \psi \lesssim 11$ degrees, while it is significantly lower when $\psi \gtrsim 11$ degrees. This maximum is not observed when the span-wise spacing is $3.49D$, because then the average power at the third row is decreasing when $\psi \gtrsim 7$ degrees due to the wakes created by turbines in the first row. In order to compare the two different span-wise spacings in more detail we plot them together in figure \ref{figure5}. This figure shows the average power ratio in the second and third row as function of the alignment angle $\psi$. Panel (a) of this figure reveals that the power output in the second row becomes equal to the power output in the first row when $\psi \gtrsim 11$ degrees, independent of the span-wise spacing. Figure \ref{figure5}b shows that for $\psi\gtrsim 7$ degrees the power ratio in the third row is significantly higher when the span-wise spacing is $5.23D$ than when the span-wise spacing is $3.49D$. Note that for the span-wise spacing of $3.49D$ the power production at the second turbine row is sometimes slightly higher than at the first row. This is due a speedup of the wind in between turbines placed on the first row and has been noted and studied in detail in prior wind tunnel experiments \cite{mct13} and simulations \cite{mct13b}.

\begin{figure}
\subfigure[$\psi=0.00$ degrees]{\includegraphics[width=0.49\textwidth]{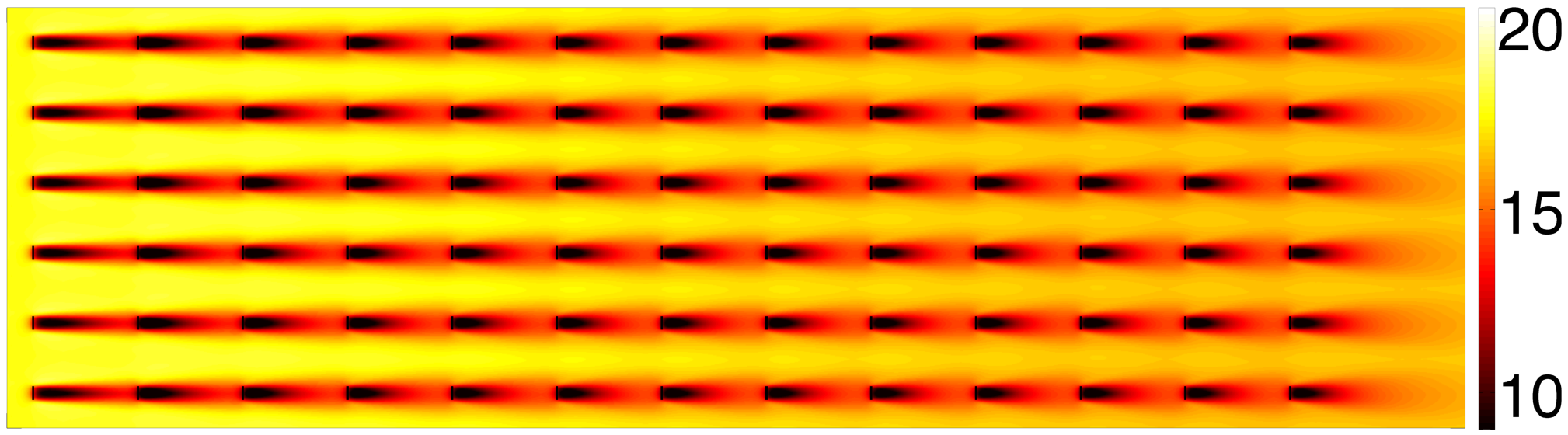}}
\subfigure[$\psi=7.59$ degrees]{\includegraphics[width=0.49\textwidth]{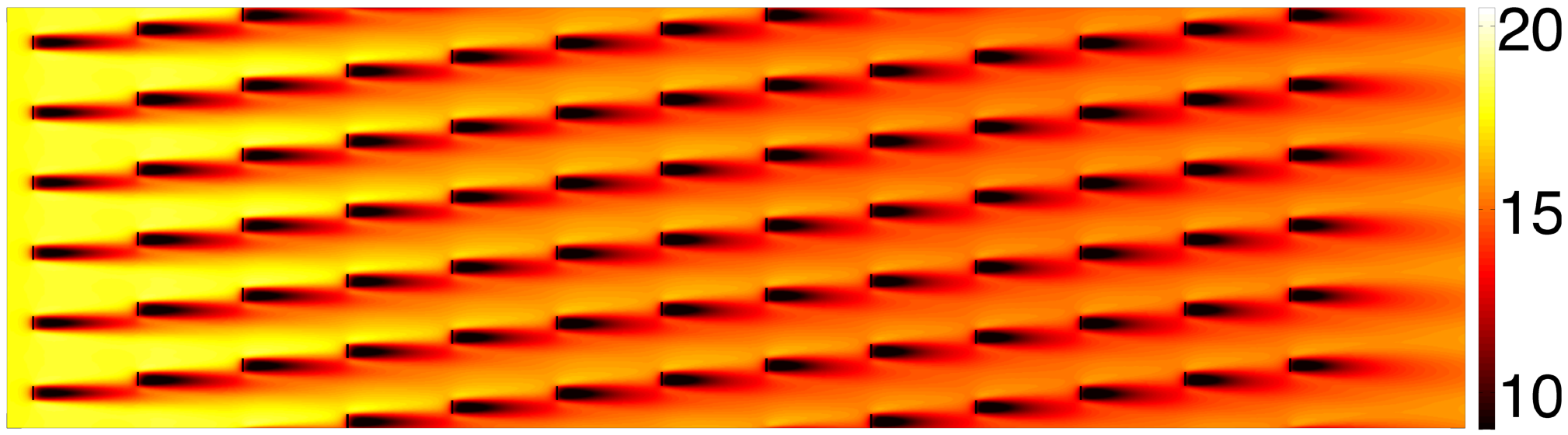}}
\subfigure[$\psi=11.31$ degrees]{\includegraphics[width=0.49\textwidth]{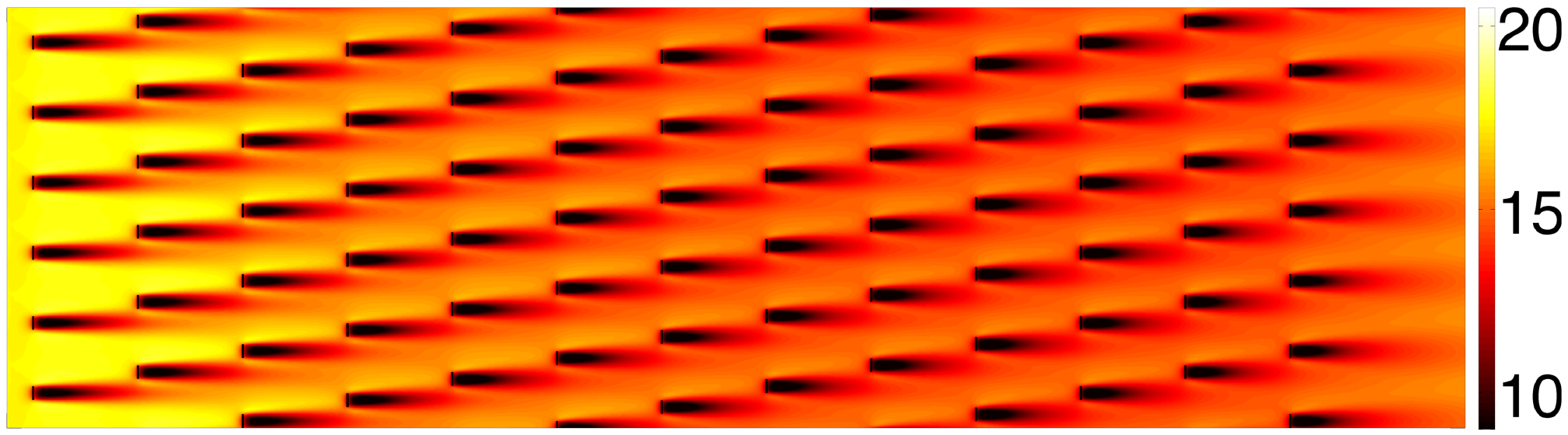}}
\subfigure[$\psi=18.43$ degrees]{\includegraphics[width=0.49\textwidth]{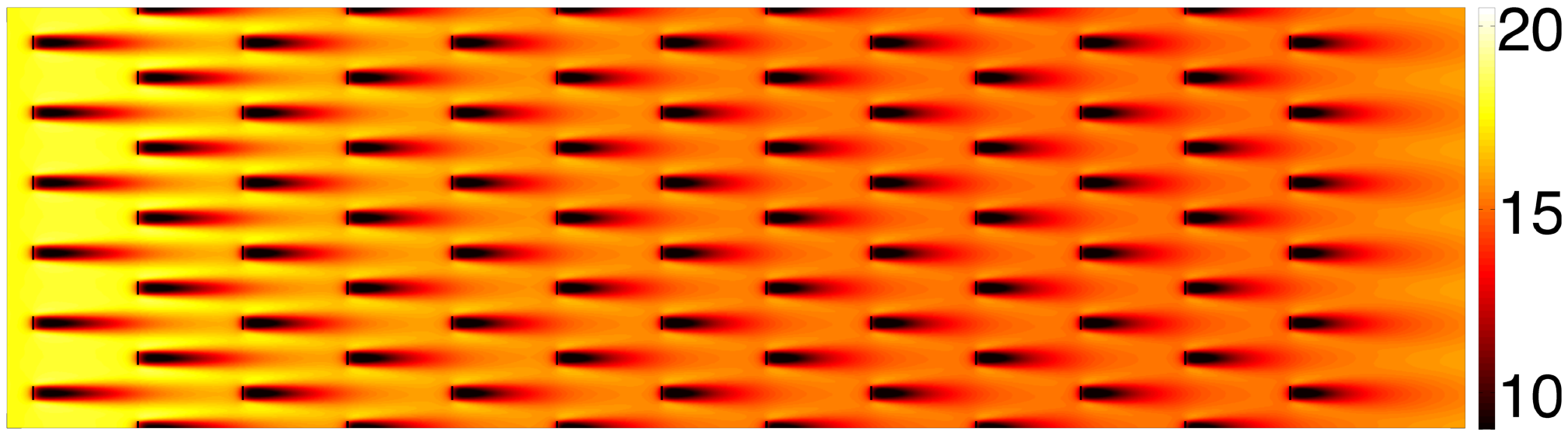}}
\caption{Time-averaged stream-wise velocity at hub-height for the wind farm with a stream-wise spacing of $7.85D$ and a span-wise spacing of $5.23D$ (case A3) for (a) $\psi=0$ degrees (aligned) (b) $\psi=7.59$ degrees (highest average power output in fourth row) (c) $\psi=11.31$ degrees (highest average power output in the third row) and (d) $\psi=18.43$ degrees (staggered). The color scale indicates the magnitude of $u/u_{*}$, which is the stream-wise wind velocity in units of friction velocity.}
\label{figure6}
\end{figure}

\begin{figure}
\subfigure[$\psi=0.00$ degrees]{\includegraphics[width=0.49\textwidth]{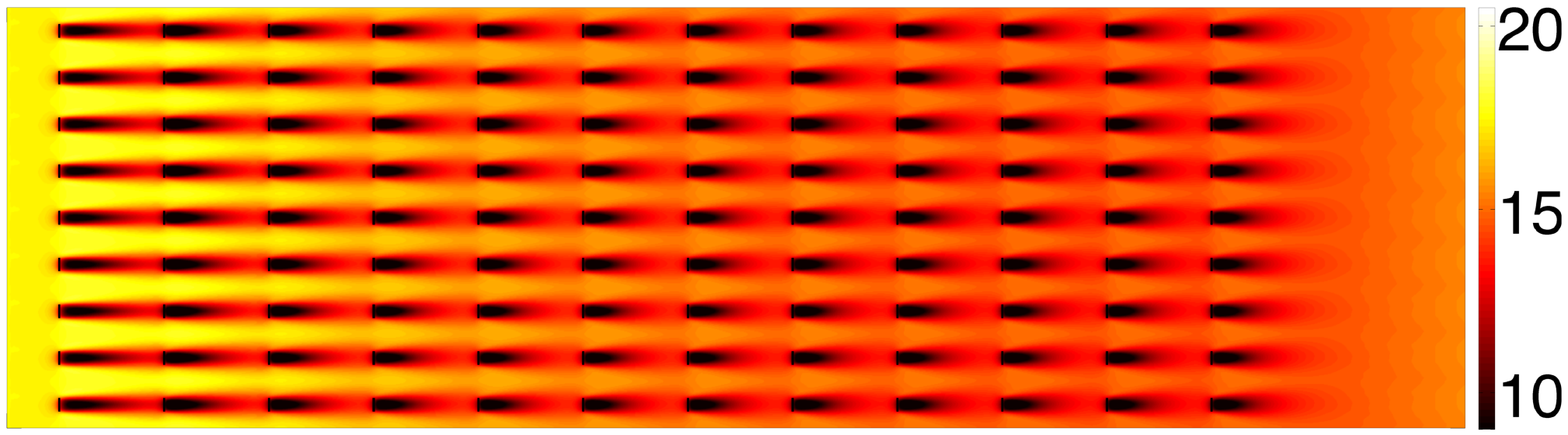}}
\subfigure[$\psi=6.77$ degrees]{\includegraphics[width=0.49\textwidth]{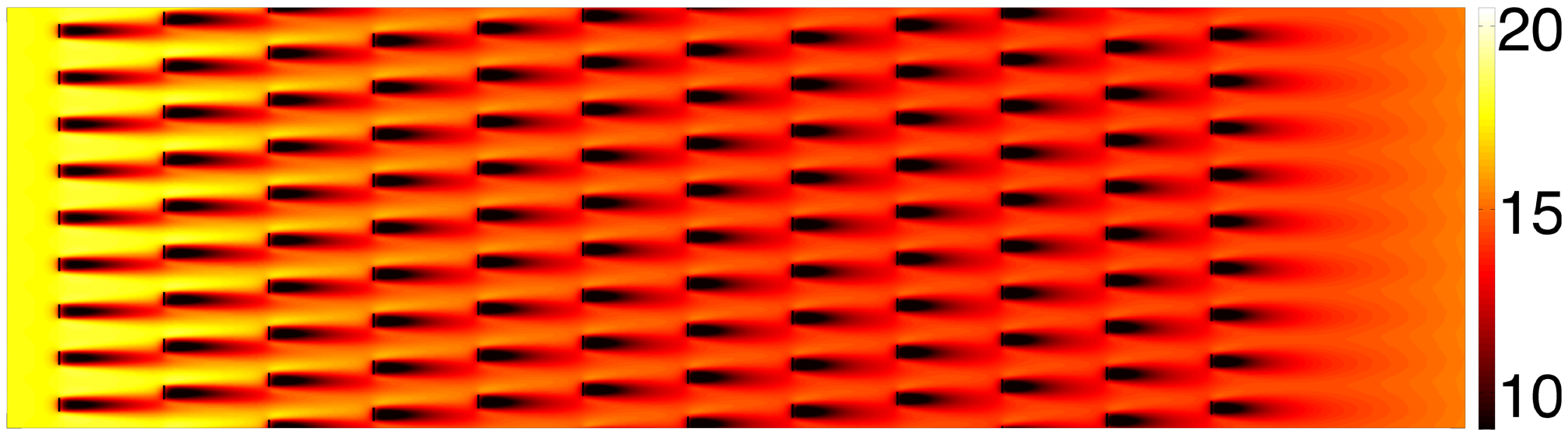}}
\subfigure[$\psi=9.46$ degrees]{\includegraphics[width=0.49\textwidth]{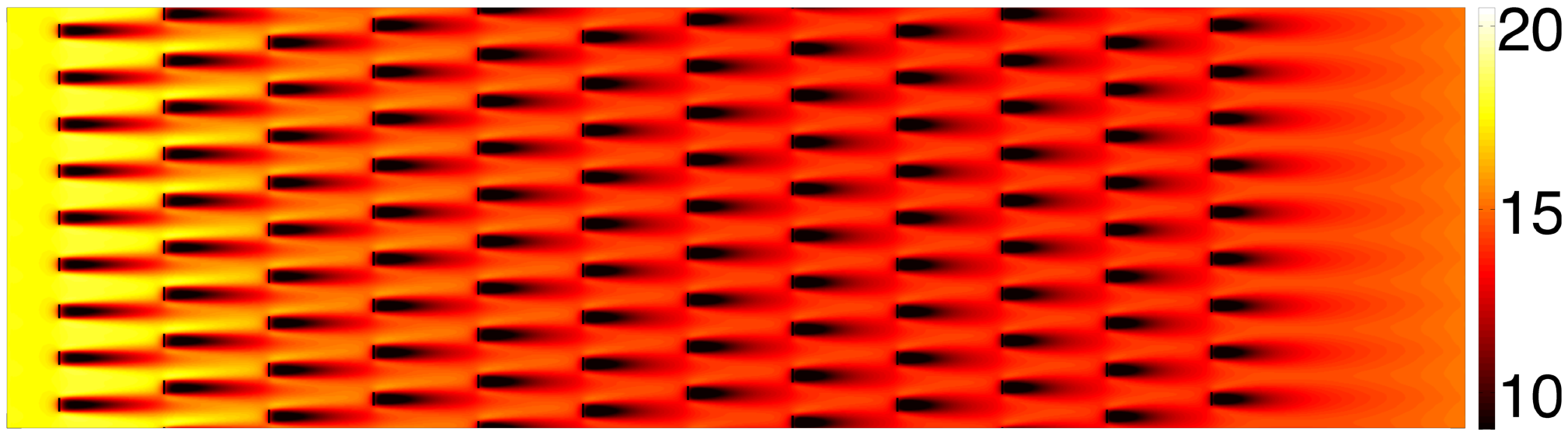}}
\subfigure[$\psi=12.57$ degrees]{\includegraphics[width=0.49\textwidth]{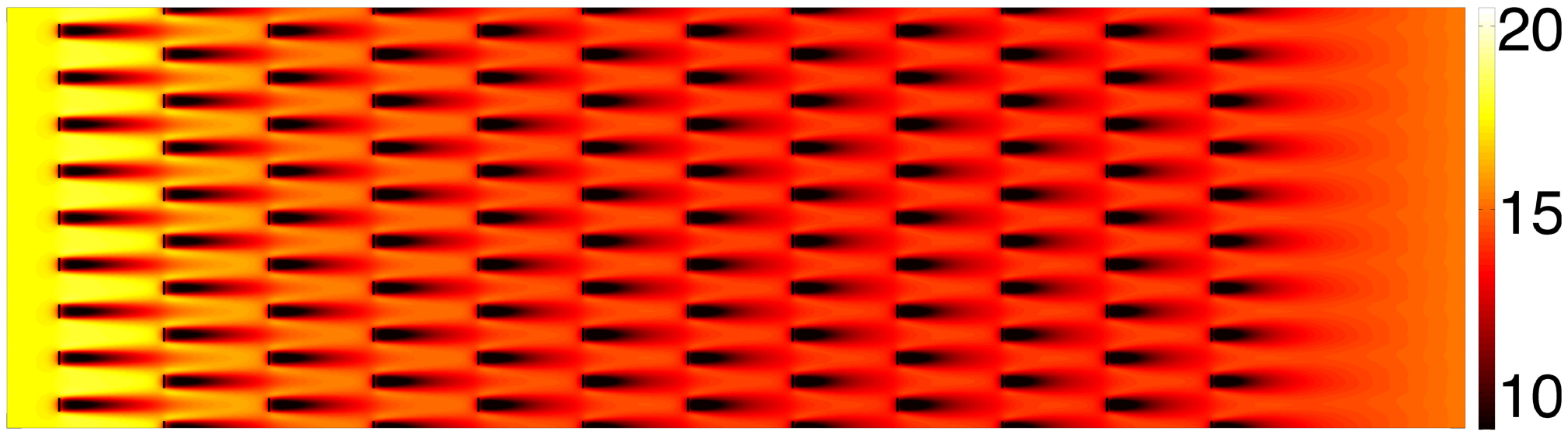}}
\caption{The time-averaged stream-wise velocity at hub-height for the wind farm with a stream-wise spacing of $7.85D$ and a span-wise spacing of $3.49D$ (case A3) for (a) $\psi=0$ degrees (aligned) (b) $\psi=6.77$ degrees (c) $\psi=9.46$ degrees and (d) $\psi=12.57$ degrees (staggered). The color scale indicates the magnitude of $u/u_{*}$, which is the stream-wise wind velocity in units of friction velocity.}
\label{figure7}
\end{figure}

\begin{figure}
\subfigure[]{\includegraphics[width=0.49\textwidth]{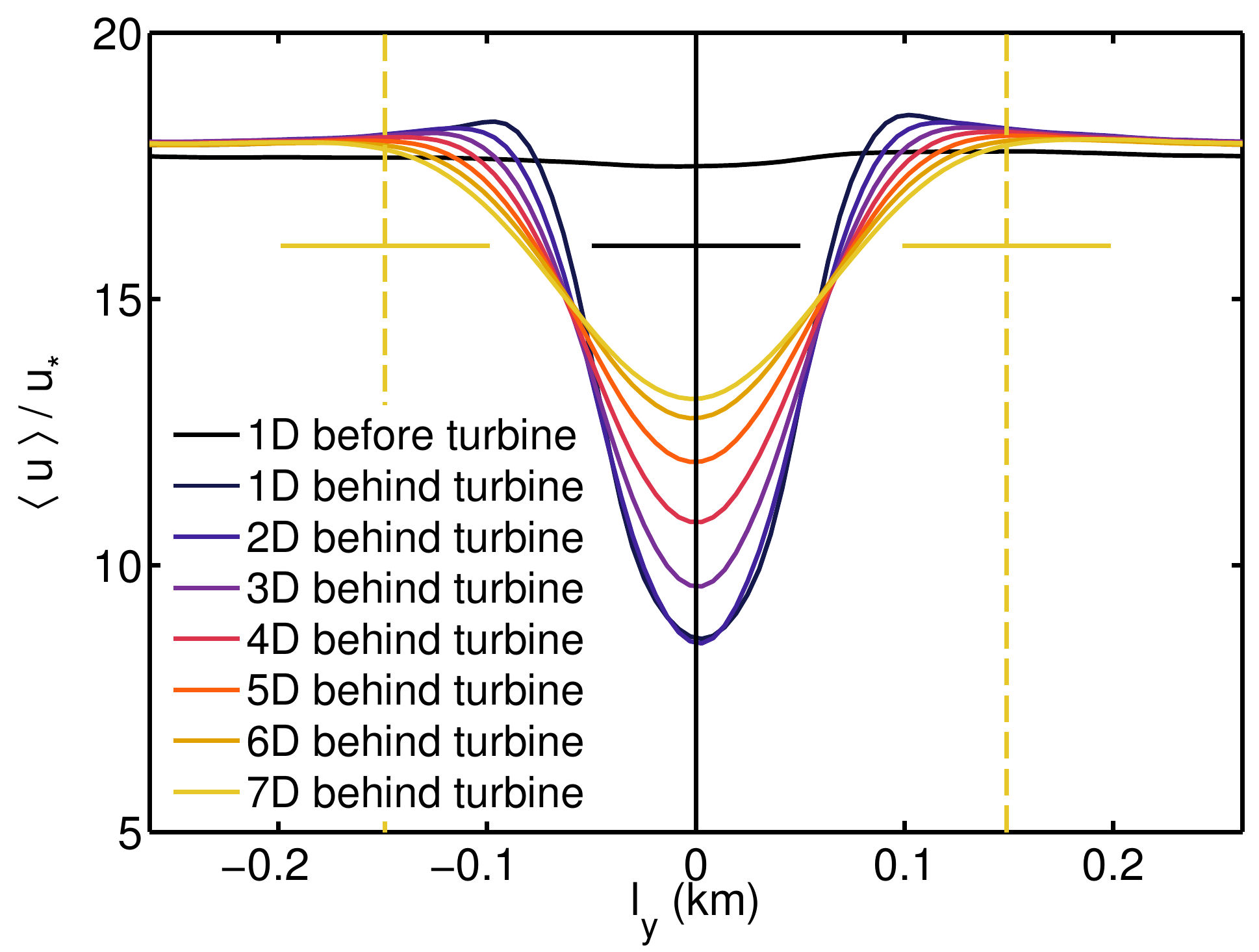}}
\subfigure[]{\includegraphics[width=0.49\textwidth]{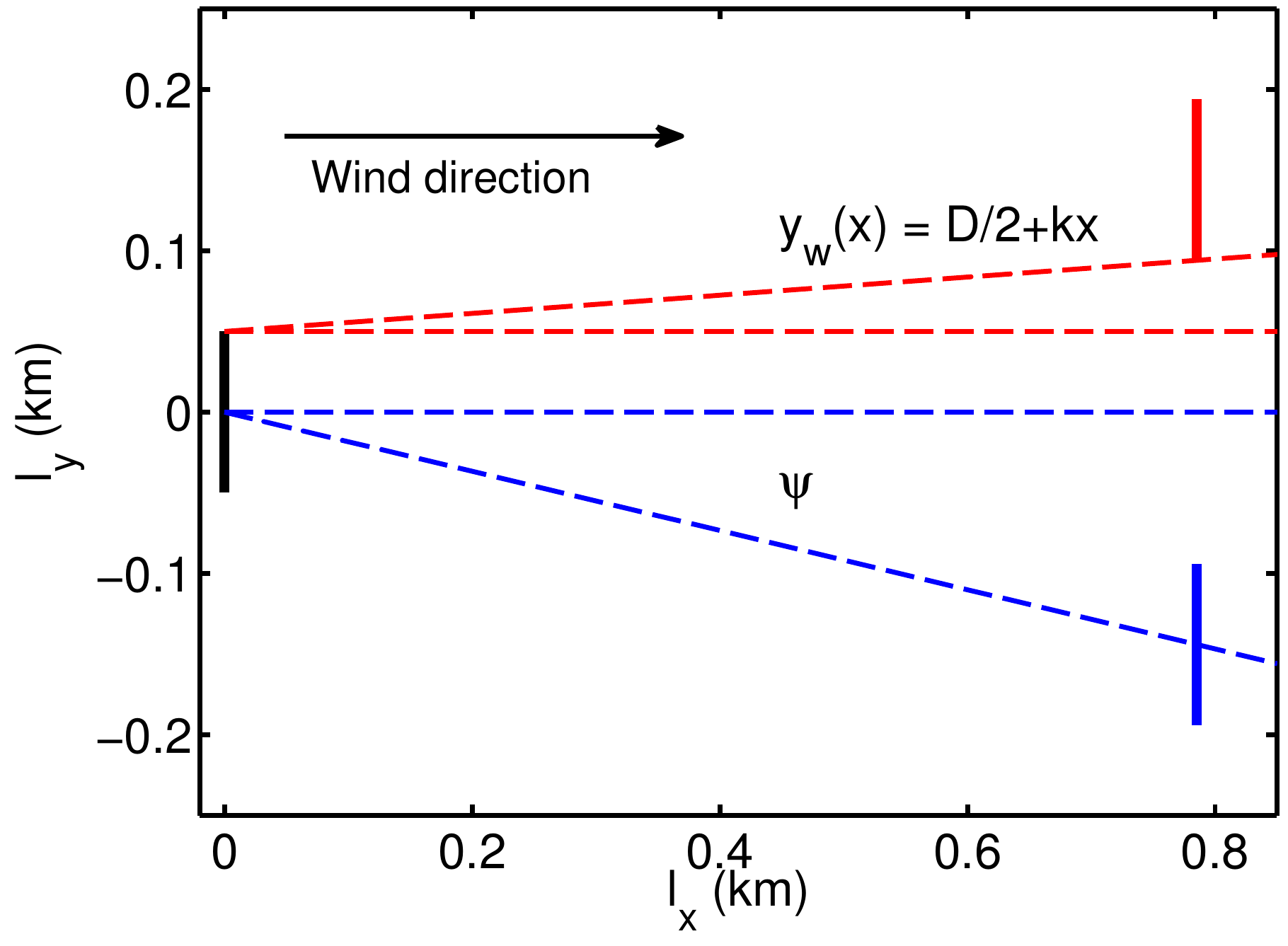}}
 \caption{(a) The average velocity profiles in front of and behind a turbine in the first row. The span-wise location of the turbine is indicated by the central black horizontal thick line, and its center by the vertical black line. Based on the average velocity profile seven diameters behind this wind turbine (approximately $1D$ in front of turbines in the second row) a turbine in the second row must be offset by $149$ m ($1.49D$) to get the same power output as the first row turbines. These locations are indicated by the dashed vertical lines and the corresponding horizontal lines, i.e. one at $l_y \approx -0.12$km and one at $l_y \approx 0.12$km, in the color corresponding to the velocity profile $7D$ behind the turbine. This spacing corresponds to an alignment angle of $\psi \approx 10.7$ degrees. (b) In this sketch of the hub-height plane the solid vertical line at $l_x=0$ km indicates the position of a turbine in the first row. The definition of the turbine alignment angle $\psi$ and the wake expansion rate k used in the model are also indicated in this sketch.}
\label{figure8}
\end{figure}

In order to study these trends in more detail we consider the time-averaged stream-wise velocity. Figure \ref{figure6} shows the results for the span-wise spacing of $5.23D$ and figure \ref{figure7} for the span-wise spacing of $3.49D$. As periodic boundary conditions in the span-wise direction are used all statistics are periodic in this direction. To improve the statistics we have therefore averaged the stream-wise velocity in $1/6^{th}$ of the original span-wise domain in figure \ref{figure6} and in $1/9^{th}$ of the original domain in figure \ref{figure7}, i.e.\ the periodicity imposed by the turbines, and subsequently this averaged velocity profile is shown over the original span-wise domain. The same procedure is used for the data shown in figure \ref{figure8}a.

We now investigate the cases with the larger span-wise spacing of $5.23D$. Figure \ref{figure8}a shows the average velocity profiles in front of and behind the turbines in the first row. This figure shows that one diameter upstream of the first turbine row the inflow profile is nearly uniform. A small reduction of the wind velocity directly in front of the turbine, due to the turbine blockage effect, is visible. The small bumps in the wake profiles just behind the turbine are due to the flow induction and has also been observed in experiments \cite{ren09}. The velocity profiles clearly reveal the effect of the wake expansion and meandering on the averaged flow velocity at hub-height. Based on the velocity profile at seven diameters behind the first turbine row, i.e. about one diameter upstream of the second turbine row, we calculate the minimal span-wise displacement that is necessary to allow the second turbine row to produce the same power as the first turbine row. This position is indicated in figure \ref{figure8}a and corresponds to an alignment angle $\psi$ of about $10.7$ degrees. This alignment angle is very close to the minimal alignment angle of $11$ degrees that our study finds to be the value at which the turbines in the second row produce the same power as turbines in the first row, see figure \ref{figure5}a. This compares well to the results in figure 8a of Hansen et al.\ \cite{han12}. This figure reveals that in Horns Rev an angle of about $11$ degrees is necessary to ensure that turbines on the second row produces the same power as turbines on the first row. One can also notice that the normalized velocity behind the turbines can sometimes be slightly larger than one. The reason is the acceleration of the wind between the turbines on the first row \cite{mct13,mct13b}, which can lead to the slightly higher power production for turbines on the second row compared to turbines on the first row. In figure \ref{figure6} and \ref{figure7} we can see that this effect is stronger when the span-wise spacing is smaller, which is in agreement with the results from \cite{mct13,mct13b}. A horizontal plane view of the downstream development of the average velocity is depicted in figure \ref{figure6} and \ref{figure7}.

Figure \ref{figure6} reveals that for $\psi \approx 11$ degrees both the turbines in the second and third row have a nearly undisturbed inflow. For smaller alignment angles $\psi$ the inflow conditions for turbines in the second and third row are influenced by the expanding and meandering wakes of the turbines directly upstream. For $\psi \gtrsim 11$ the turbines in the third row are subjected to the wake created by the first turbine row. In fact, for the staggered configuration ($18.43$ degrees) the turbines in the third row are directly in the wakes created by the turbines in the first row, both of these scenarios limit their power production. Therefore we conclude that $\psi_*$, i.e. the $\phi$ with the highest average power output, is obtained when the first turbine falls just outside of the wake of the nearest upstream turbine. 

\begin{figure}
\subfigure{\includegraphics[width=0.49\textwidth]{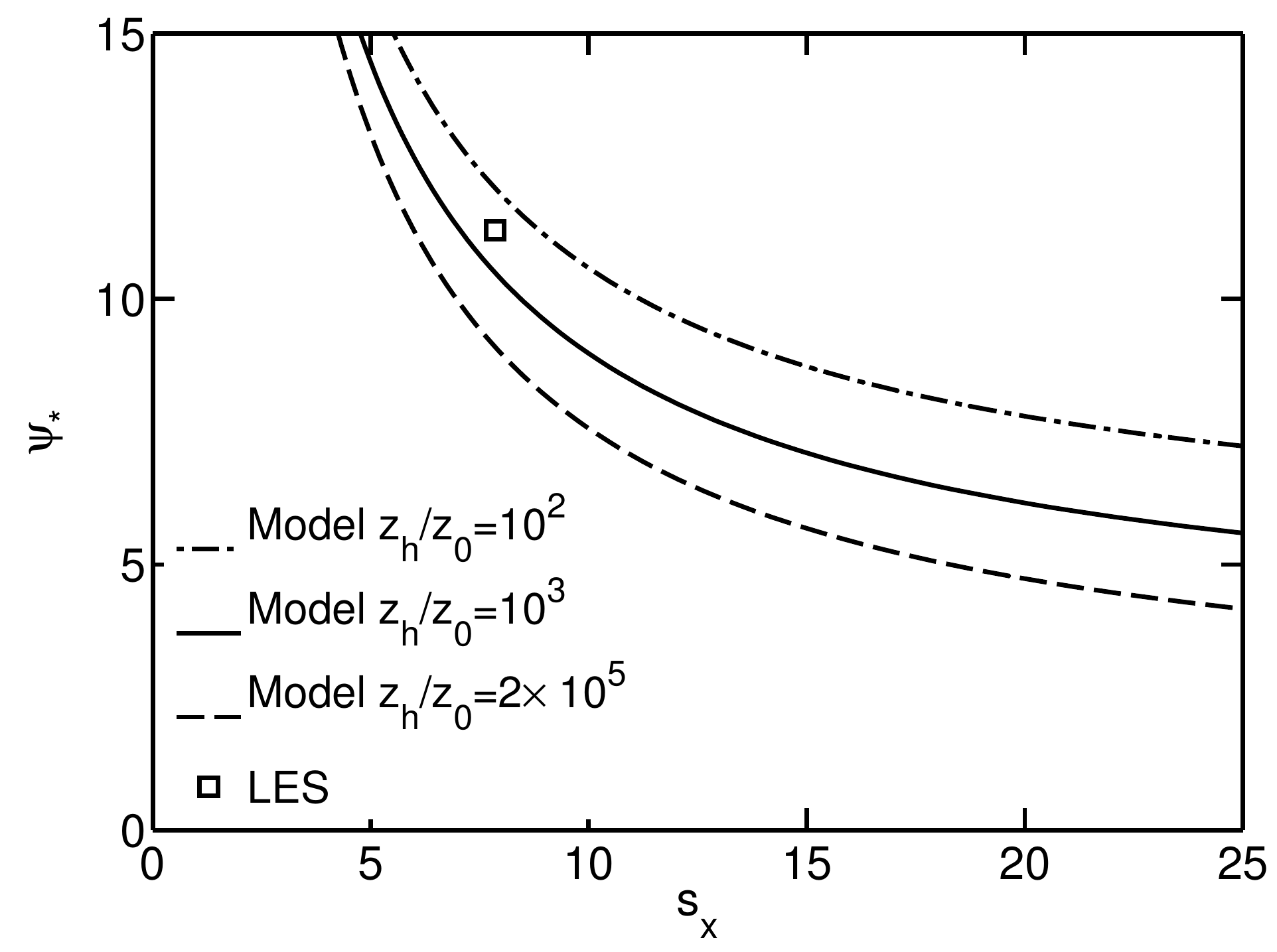}}
 \caption{$\psi_{*}$ as function of $s_x$ based on model equation (\ref{angle_model}) with $s_x=7.85$ and $z_h/z_0=10^{3}$. The results for two different $z_h/z_0$ combinations are given for reference.}
\label{figure9}
\end{figure}

In fact $\psi_{*}$ can be related to wake expansion rates commonly used in wake models \cite{jen84,kat86} as follows. We assume a wake expansion rate of $k$, the edge of the wake follows $y_w(x) = D/2+k x$, where $x$ is the distance to the upstream turbine, see figure \ref{figure8}b. When $S_x=s_x D$ the edge of the wake is at $D(1/2+k s_x)$ and the next turbine should be displaced by an additional $D/2$, i.e.\ we wish to have $S_{dy}=D(1+k s_x)$. As a result, the angle obeys
\begin{equation}
	\psi_{*} = \arctan \left(\frac{[1+k s_x]D}{s_xD}\right) = \arctan\left(k+1/s_x \right).
\end{equation}
A common estimate of the lateral growth rate of wakes is the ratio of the lateral fluctuating velocity and the mean advective velocity at hub-height, i.e. $k = {u_*}/{\overline{u}_h}.$ As the placement of turbines on the second row with respect to the turbines on the first row is most important we use the upstream flow characteristics, i.e. 
\begin{equation}
\frac{\overline{u}}{u_*} = \frac{1}{\kappa} \ln\left(\frac{z}{z_0}\right)
\end{equation}
to derive that
\begin{equation}
\label{angle_model}
	 \psi_{*} = \arctan\left( \frac{\kappa}{\ln\left(\frac{z_h}{z_0}\right)} +\frac{1}{s_x} \right).
\end{equation}
For our case $s_x=7.85$, $z_h/z_0=1000$, and $\kappa=0.4$. Figure \ref{figure9} shows that the modeled $\psi_{*}$ as function of $s_x$ agrees reasonably well with the LES result. Note that $\lim_{s \to 0} \psi_{*} = 90^{\circ}$, which corresponds to turbines placed next to each other, and $\lim_{s \to \infty} \psi_{*} = \arctan (\kappa/\ln(z_h/z_0)) \approx 3.3^{\circ}$, which corresponds to the assumed wake expansion in the far field.The results for two additional zh/z0 values are shown in figure 9, where it is clear that zh/z0 has an effect on the predicted $\psi_*$. Here we note that for $z_h=100$m, $z_h/z_0 = 10^2$ would correspond to $z_0=1$m, loosely representing a forested region, see Ref.\ \cite{wie01}, while, $z_h/z_0 = 10^3$ corresponds to $z_0=10$cm, i.e. a land surface covered with shrubs, while $z_h/z_0= 2\times10^5$ corresponds to $z_0=2$mm, which is loosely representative of conditions for fairly rough ocean waves \cite{wie01}.

For the smaller span-wise spacing the average velocity profiles in figure \ref{figure7} show that the wakes created by the first row reach the third turbine row when $\psi \gtrsim 7$ degrees. This effect is also clearly observed in figure \ref{figure5}b, which shows that for both span-wise spacings, the average power output at the third row is the same when $\psi \lesssim 7$ degrees. However, a significantly higher average power output is obtained at the third row when $\psi \gtrsim 7$ degrees and the span-wise spacing is larger. The reason is that the larger span-wise spacing prevents the turbines in the third row from being influenced by the wakes created by the turbines in the first row, while this is not the case for the smaller span-wise spacing. 

\begin{figure}
\subfigure[]{\includegraphics[width=0.49\textwidth]{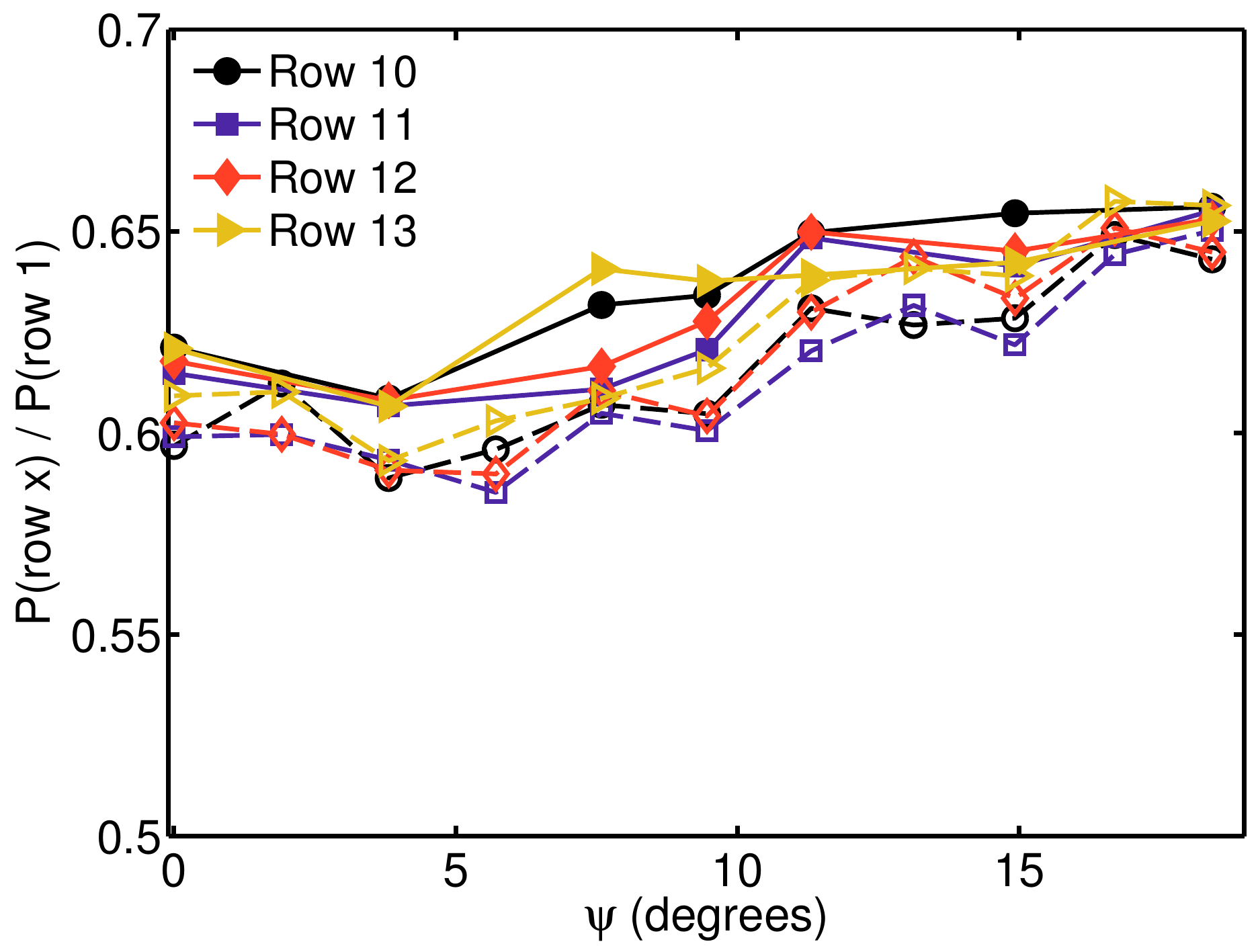}}
\subfigure[]{\includegraphics[width=0.49\textwidth]{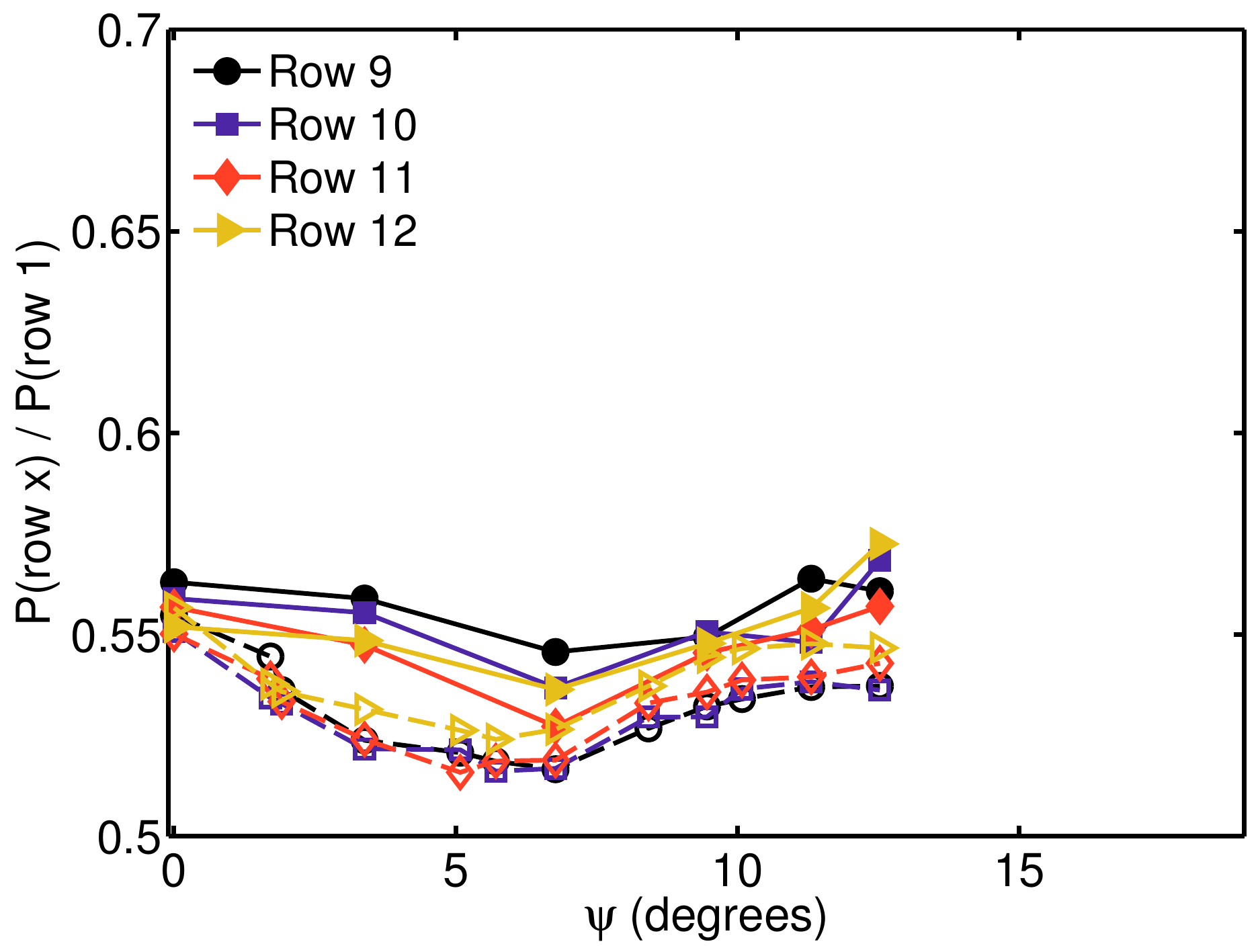}}
\caption{The average power ratio in the fully developed regime of the wind farm for the larger ($5.23D$) and smaller ($3.49D$) span-wise spacing is shown in panels (a) and (b). The open and closed symbols indicate the results on the coarsest (cases *1) and the finest resolution (cases *3). }
\label{figure10}
\end{figure}

\begin{figure}
\subfigure{\includegraphics[width=0.49\textwidth]{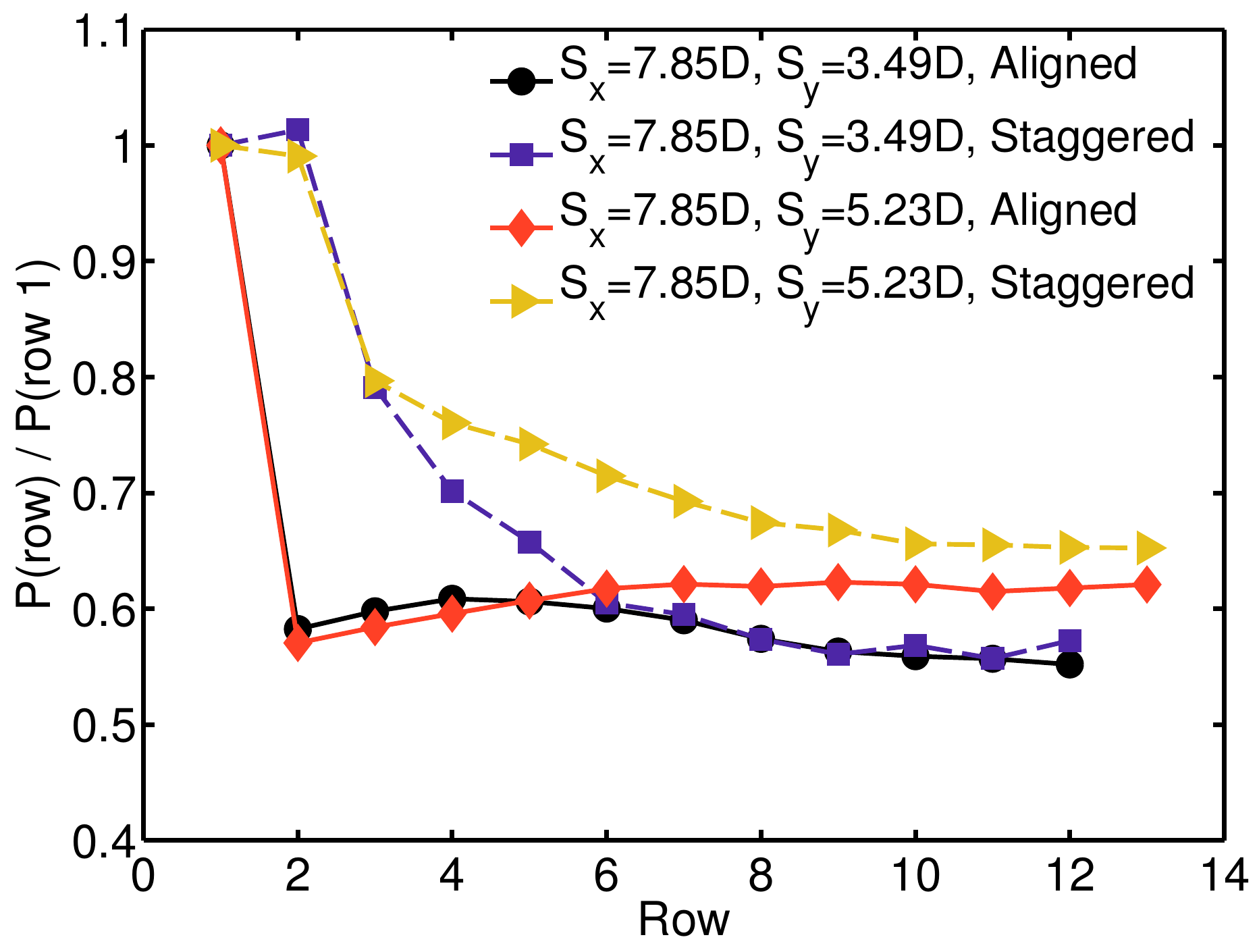}}
\caption{The normalized power output based on the power output of the first row as function of the downstream position for aligned and staggered wind farms.For a definition of the case numbers see table \ref{table1}.}
\label{figure11}
\end{figure}

Figure \ref{figure4}b shows that for the fifth and subsequent rows a small further reduction in the average turbine power output is observed when the span-wise spacing is $5.23D$ and $\psi>0$. The reason is that the average kinetic energy that is available at hub-height slowly decreases further downstream from the first turbine row. The reduction in kinetic energy slowly reduces the wake recovery rate, as less energy can come from the sides, until the fully developed regime is reached. Such behavior is difficult to capture in traditional wake models for wind farm design. This gradual reduction of the power output as function of the downstream position is not visible when the span-wise spacing is $3.49D$, see figure \ref{figure4}d. In that case the higher turbine density ensures that most of the available kinetic energy in the hub-height plane is already extracted from the hub-height plane at the fifth and sixth turbine rows.

In the fully developed regime the vertical kinetic energy flux that is created by the turbine wakes balances the average power that is extracted by the turbines \cite{cal10,cal10b} and in figures \ref{figure4}b and \ref{figure4}d the average turbine power output seems nearly independent of the angle $\psi$. Nevertheless some small differences can be seen in the fully developed regime of the wind farm. These differences can be attributed to the uncertainty due to limited time averaging and the grid resolution. Figure \ref{figure10} shows that for a span-wise spacing of $5.23D$ the average power output in the fully developed regime is higher for a staggered configuration than for an aligned configuration, which is in agreement with the observations of Wu and Port\'e-Agel \cite{wu13}. This difference between the aligned and staggered configuration is not observed for the smaller span-wise distance. This effect can also observed in figure \ref{figure11}, which compares the average power output for the aligned and staggered configurations as function of the downstream position.

\begin{figure}
\subfigure[]{\includegraphics[width=0.49\textwidth]{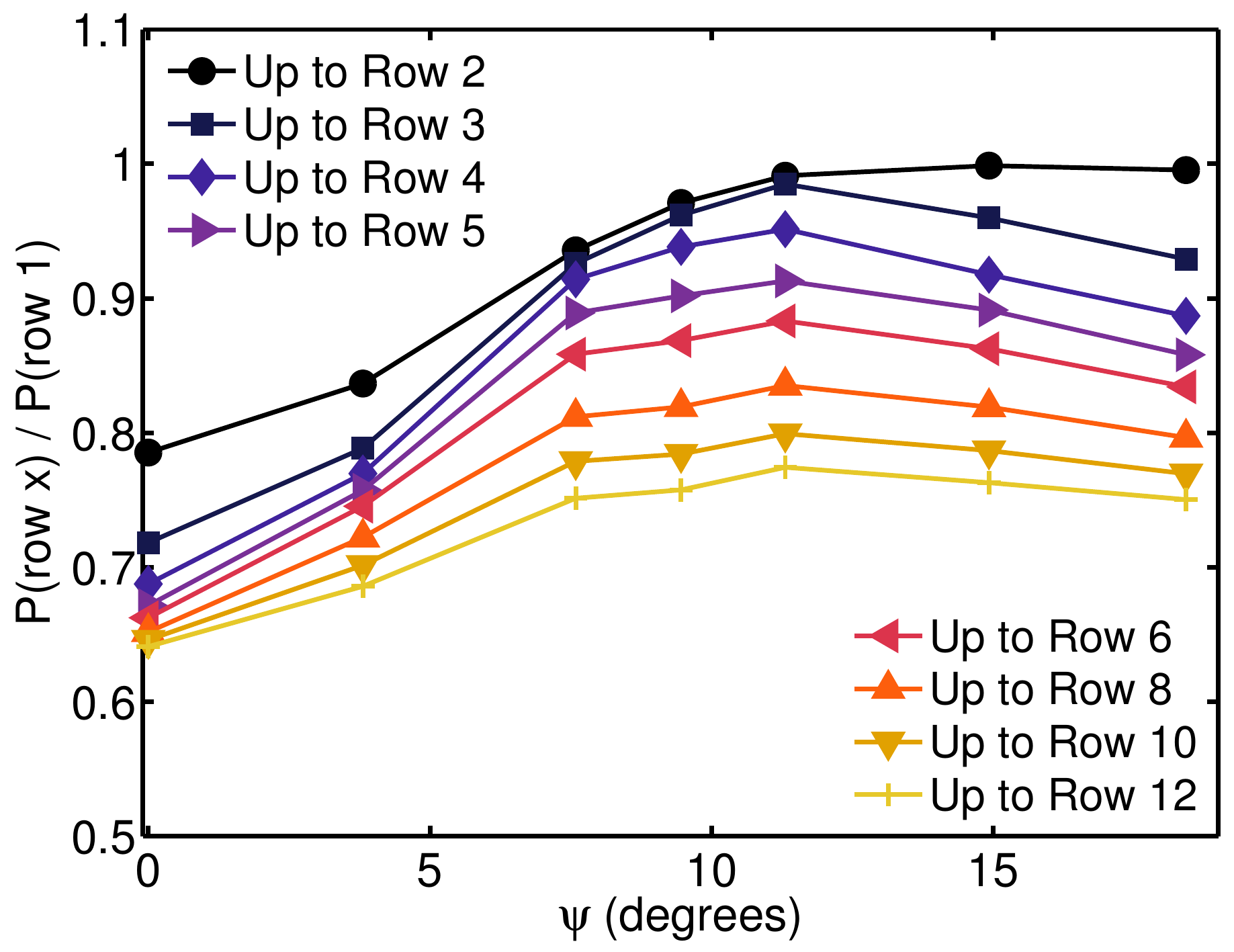}}
\subfigure[]{\includegraphics[width=0.49\textwidth]{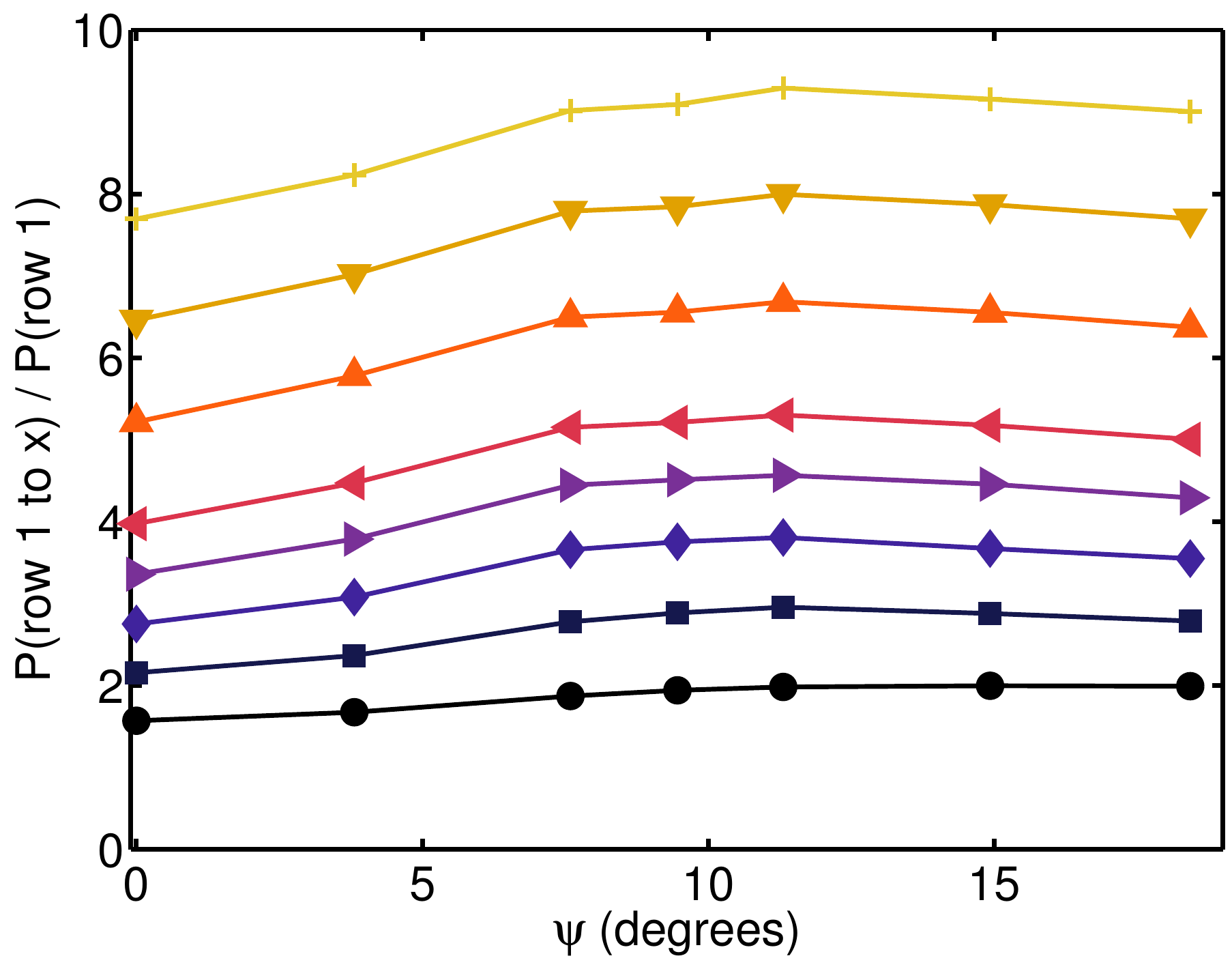}}
\subfigure[]{\includegraphics[width=0.49\textwidth]{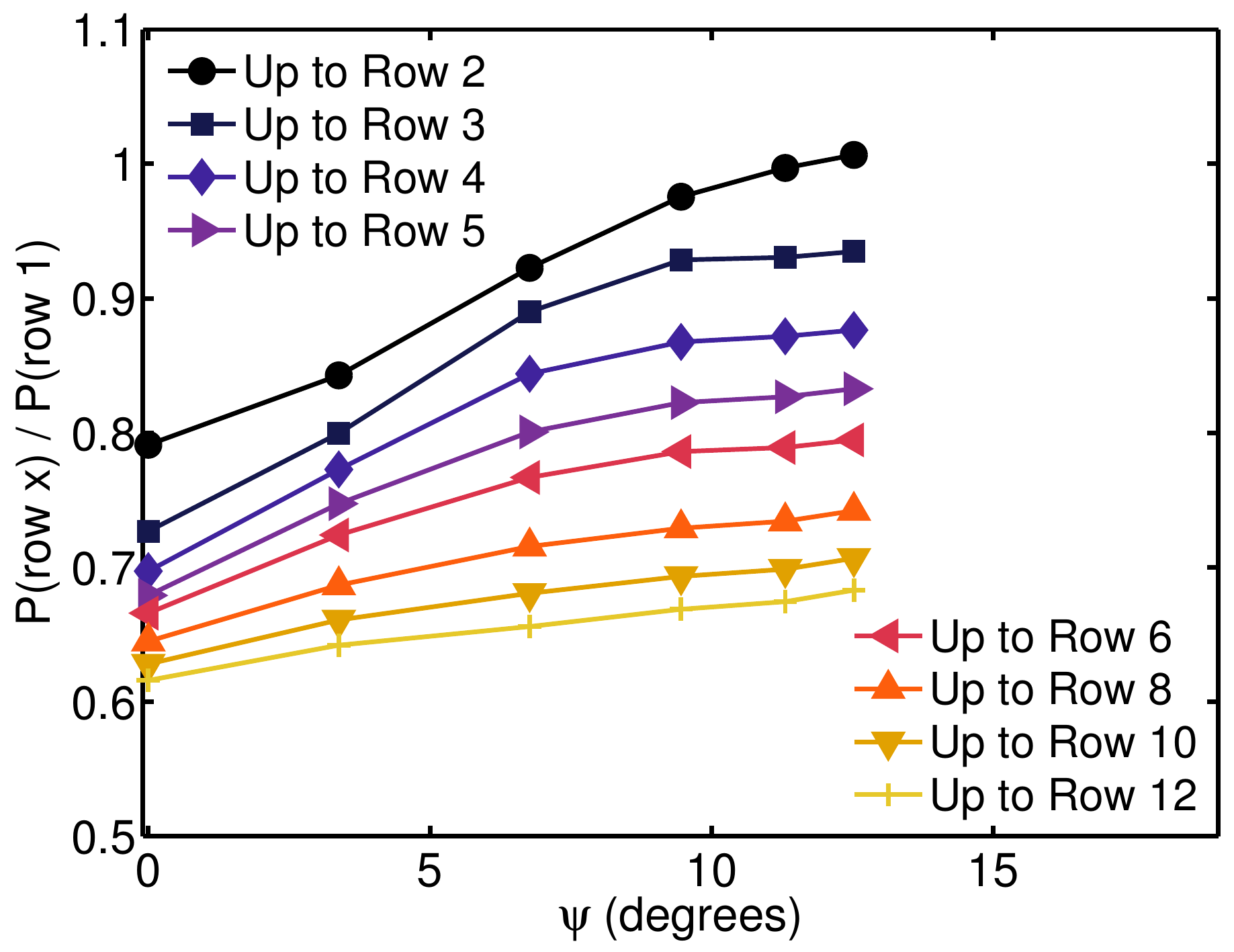}}
\subfigure[]{\includegraphics[width=0.49\textwidth]{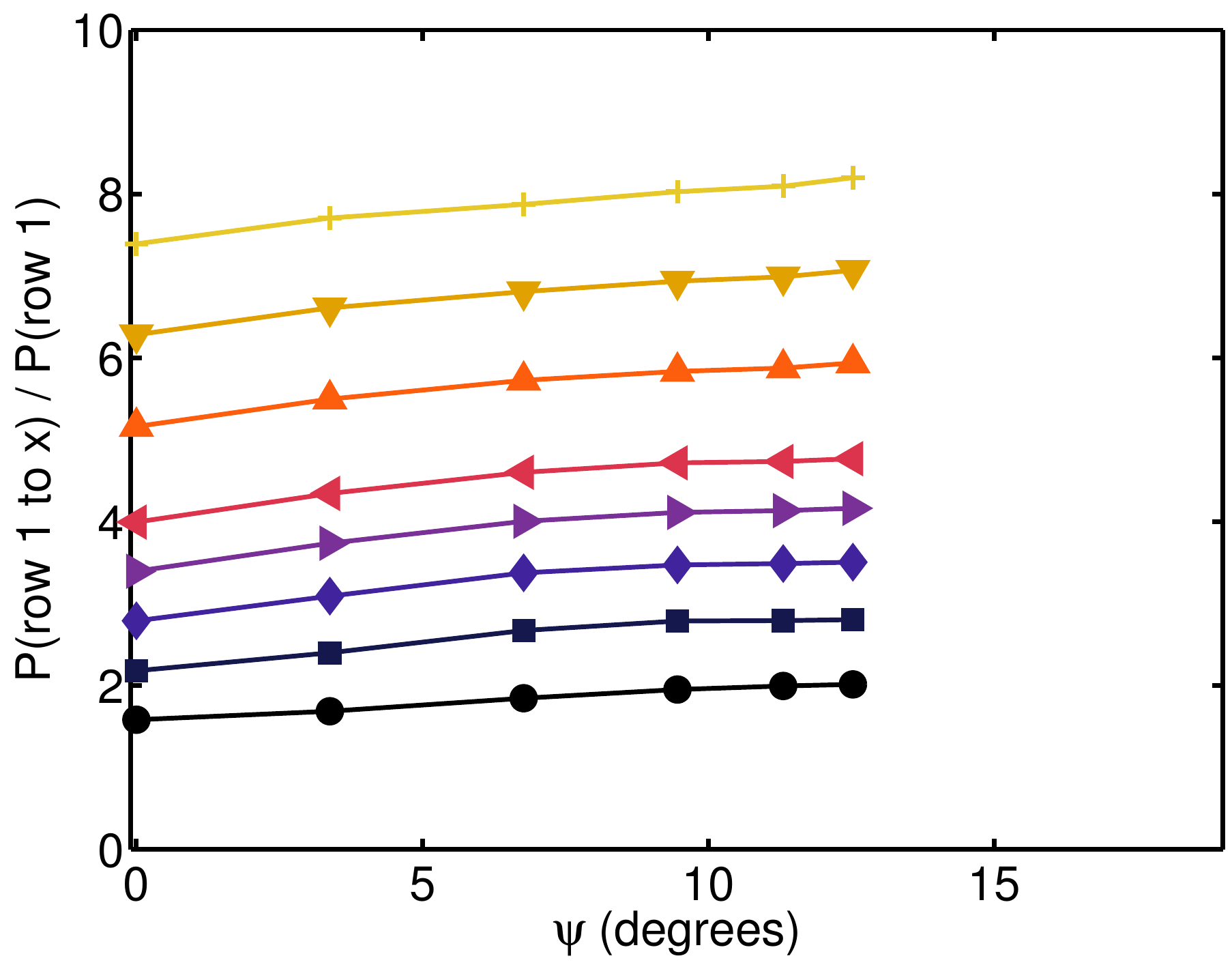}}
 \caption{The top (a,b) and bottom (c,d) panels indicate average turbine power ratios for the wind farm with a span-wise spacing of $5.23D$ (case A3) and $3.49D$ (case B3), respectively. The left panels (a and c) show results for different wind farm lengths as function of the alignment angles $\psi$. The corresponding right panels (b and d) show the total average power output for the different cases normalized by the average power output of turbines in the first row.}
\label{figure12}
\end{figure}

The results discussed above confirm that the average power output that is found at a particular turbine row depends significantly on the alignment with the incoming flow (or span-wise shift between rows). For the configurations studied here, we find that for a span-wise spacing of $5.23D$ the highest average power output for the third row is obtained with an alignment angle of $\psi \approx 11$ degrees, for the fourth row for $\psi \approx 8$ degrees, and in the fully developed regime for the staggered configuration ($\psi=18.43$ degrees). Therefore one may wonder which wind farm layout gives the highest average power output for the entire wind farm. Assuming that the average power output of turbines is not influenced by downstream turbines we can calculate the average power output per turbine for different wind farm lengths. Figure \ref{figure12}a shows the average turbine power output normalized by the average power produced by turbines in the first row for wind farms of different lengths. For the larger span-wise spacing of $5.23D$ the highest average power output is obtained for $\psi\approx 11$ degrees. Figure \ref{figure12}a shows that for a wind farm with $12$ rows in the stream-wise direction the average turbine power output can range from approximately $60\%$ to $75\%$ of the average power produced by turbines in the first row. Figure \ref{figure12}c shows that for the smaller span-wise spacing of $3.49D$ the highest average power output is also obtained for alignment angles of $\psi \approx 11$ degrees, which for this span-wise spacing is a nearly staggered orientation. Although an alignment angle of about 7 degrees gives the highest power output at the third row of a wind farm with the 3.49D span-wise spacing the total power produced over the wind farm is highest when $\psi \approx 11$ degrees. This can be explained by noticing that there is a much higher power output in the second row for the $11$ degree orientation.

\begin{figure}
\subfigure[]{\includegraphics[width=0.49\textwidth]{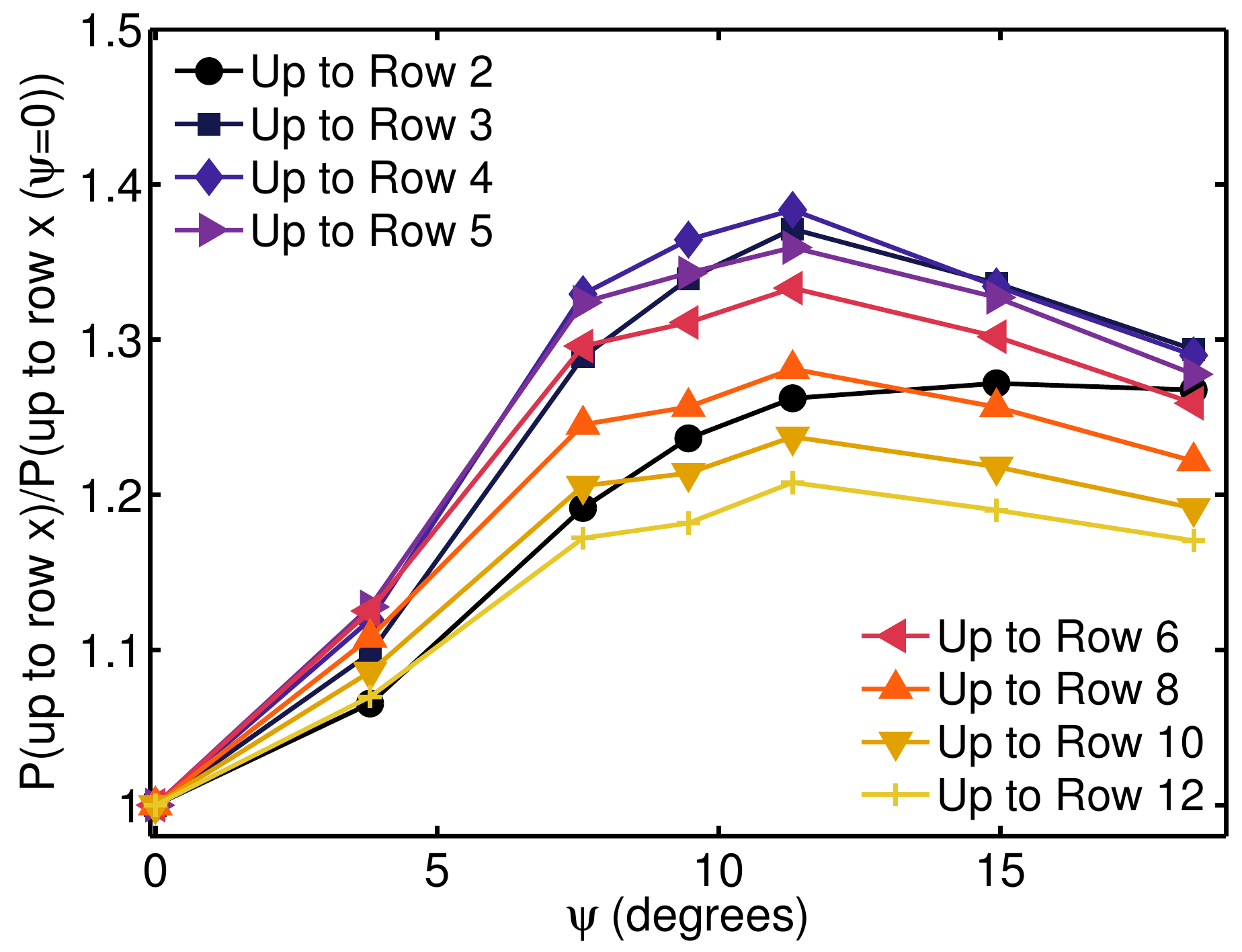}}
\subfigure[]{\includegraphics[width=0.49\textwidth]{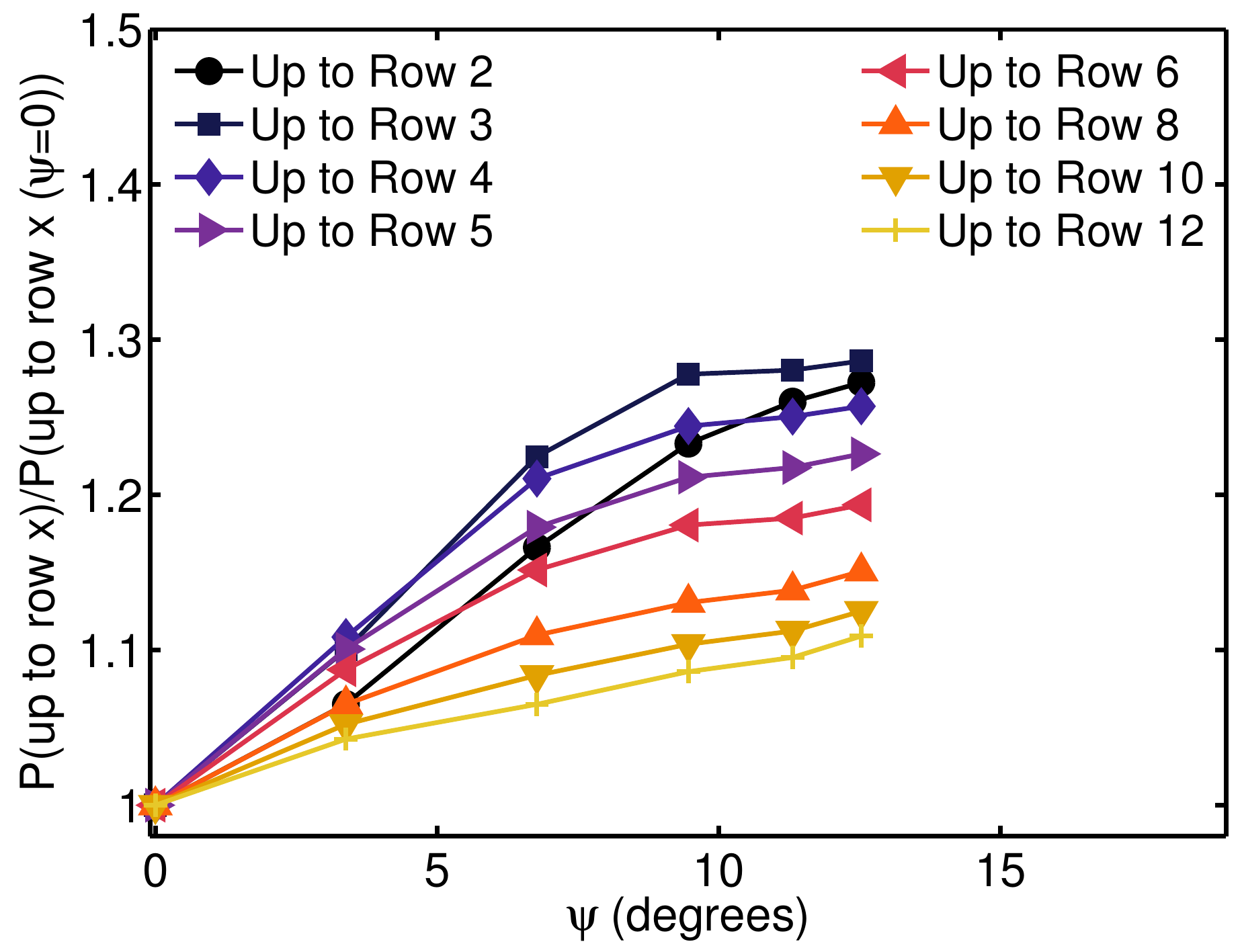}}
 \caption{Panel a and b show average power production for different with farm lengths normalized by the corresponding production of an aligned wind farm as function of the alignment angle $\psi$ for the larger ($5.23D$, case A3) and smaller ($3.49D$, case B3) span-wise spacing, respectively.}
\label{figure13}
\end{figure}

Figure \ref{figure13} shows the average power turbine power output for different wind farm lengths normalized by the average power output obtained in the aligned configuration. This figure shows that the alignment is most important for wind farms with three or four turbine rows. For shorter wind farms the relative increase is lower as a smaller percentage of the wind turbines is affected by wake effects. For longer wind farms, the relative power-output increase that can be obtained with respect to the aligned cases decreases. This effect arises because for long wind farms the average power output in rows further downstream depends less on the orientation than in the first few rows. Note that the additional power output that can be obtained by changing the wind farm configuration is larger when the span-wise distance between the turbines is larger. Figure \ref{figure13}a shows that an increase of the average power production with respect to the aligned configuration of approximately $35\%$ can be obtained when the stream-wise spacing is $5.23D$, whereas this gain is about $25\%$ when the span-wise spacing is $3.49D$.

\section{Conclusions}
In this study we have used LES to quantify relative power output in wind farms and the effects of various alignments between downstream rows of turbines. Specifically, we studied the effect of the span-wise offsets and corresponding alignment angles $\psi$ with respect to the incoming flow (see figure \ref{figure1}) on the average power output. We used a constant stream-wise spacing between the wind-turbines of $7.85D$, where $D$ is the diameter of the turbines, and span-wise spacings of $5.23D$ and $3.49D$. We found that the highest power output for the entire wind farm is obtained when an alignment angle $\psi$ of approximately $11$ degrees is used. The data indicate that the highest power output for the entire wind farm is obtained for the smallest alignment angle for which wake effects from turbines in the first few upstream rows are minimal, i.e. the minimum angle $\psi$ that ensures that the power output of the second row is close to the power output of the first row. For the smaller span-wise spacing of $3.49D$ a $\psi_*$ of $\approx 11$ degrees corresponds to a nearly staggered wind farm, while it corresponds to an intermediate alignment for the larger span-wise spacing of $5.23D$. The highest average power output for a large wind farm is thus not necessarily obtained for a fully staggered (checker-board) configuration, but can occur for intermediate off-sets and/or wind direction so that the first several rows, not just the second, fall outside wakes from the upstream row(s) of turbines. It should be noted that this work does not claim to provide or propose an optimal layout of wind farms in general. Determining the optimal layout of an actual wind farm depends critically upon annual distributions of wind alignment, site-specific wind roses, and is beyond the scope of the present work. In future work we plan to analyze LES results to also study the development and mechanisms associated with the vertical kinetic energy flux, which is crucial for the wind farm performance especially in the fully developed regime \cite{cal10}.

{\it Acknowledgements}: The authors thank Jason Graham and Claire Verhulst for interesting conversations and comments. This work is funded in part by the research program `Fellowships for Young Energy Scientists' (YES!) of the Foundation for Fundamental Research on Matter (FOM) supported by the Netherlands Organization for Scientific Research (NWO), and in part by the US National Science Foundation, grants numbers CBET 1133800 and OISE 1243482. This work used the Extreme Science and Engineering Discovery Environment (XSEDE), which is supported by the National Science Foundation grant number OCI-1053575 and the LISA cluster of SURFsara in the Netherlands.

\end{document}